\documentclass[journal ]{new-aiaa}
\usepackage[utf8]{inputenc}
\usepackage{textcomp}
\usepackage{subcaption}

\usepackage{graphicx}
\usepackage{amsmath}
\usepackage[version=4]{mhchem}
\usepackage{siunitx}
\usepackage{longtable,tabularx}
\title{Aerodynamic consequences of wing damage in dragonflies}

\author{Peng Yu \footnote{peng.yu@espci.fr}, Ramiro Godoy-Diana and Benjamin Thiria}
\affil{Laboratoire de Physique et Mécanique des Milieux Hétérogènes (PMMH), CNRS UMR 7636,
ESPCI Paris-Université PSL, Sorbonne Université, Université Paris Cité, Paris, France}
\author{Dmitry Kolomenskiy}
\affil{Skoltech Center for Design, Manufacturing and Materials, Skolkovo Institute of Science and Technology, Moscow, Russian Federation}
\author{Thomas Engels\footnote{thomas.engels@univ-amu.fr}}
\affil{CNRS \& Aix-Marseille Université, UMR 7287, Institut des Sciences du Mouvement Etienne-Jules Marey, Marseille, France}

\begin{document}

\maketitle

\begin{abstract}
Flapping wings are the primary means by which dragonflies generate forces, but they are susceptible to damage due to their inherent fragility. The damage results in a reduction in wing area and a distortion of the original wing, which in turn leads to a decline in flight ability. Furthermore, the flows of dragonfly fore- and hindwings exhibit an interaction, thus damage to the forewing can also impact the aerodynamic performance of the ipsilateral hindwing. In this study, we examine this problem through CFD (computational fluid dynamics) simulations on a series of damaged dragonfly fore-/hindwing models according to the probability of area loss from the literature. The flow fields and aerodynamic forces for the different damaged wing cases are compared with those for the intact wings. This comparative analysis reveals how the different patterns of wing damage modify the vortex structures around the flapping wings and lead to a drop in aerodynamic force production. The causes behind the diminishing aerodynamic performance are shown to be subtler than the pure area loss and are regulated by the changes in the flow field that result from wing damage. Wing-wing interaction becomes particularly important when forewing damage occurs.
\end{abstract}

\section{Introduction}
Insects generate the forces to execute flight manoeuvres with their flapping wings. This constitutes a complex aerodynamical problem in particular because of the perpetual cycle of acceleration and deceleration brought by the back and forth motion of the wing. The resulting flow around the wings is very different from that around a fixed wing gliding through the air and a vast literature has been dedicated to the understanding of the associated unsteady aerodynamic mechanisms ---see e.g. \cite{Sane:2003,Wang:2005,Sun:2014,Chin:2016}, for a review. Adding to this complexity, wings are often damaged by predators or undergo wear over their lifespan \cite{rajabiInsectWingDamage2020,fosterWhatCausesWing2011}. 
It is well known that wing damage affects flight performance \cite{jantzenHindwingsAreUnnecessary2008} and consequently the success when catching prey or escaping predators \cite{combesDynamicsAnimalMovement2010}, two factors that significantly impact biological fitness. 
Although several studies have shown how dragonfly wings break \cite{rajabiInsectWingDamage2020,rudolfFractureResistanceDragonfly2019} and where this damage is most likely to occur \cite{rajabiProbabilityWingDamage2017}, the aerodynamic aspect of dragonfly wings damage is not yet understood. In this study, we combine previously published experimental data on damage probability with CFD (Computational Fluid Dynamics) to answer the question: What is the difference in aerodynamic performance if the dragonfly wings are damaged? Consequently, we do not investigate how to adjust the wing beat kinematics in order to \emph{compensate} for this loss, but rather keep the wing beat kinematics constant. From the resulting loss in force production, we can quantify the importance of the lost area and put it in relation to the probability measured in the field, in order to evaluate the flying capacity of a dragonfly with damaged wings.

The aerodynamic force of the insect wings is generated by the work of flapping motion against air resistance. During this motion, the effective wing area is a crucial factor in force generation with different regions contributing variably \cite{engelsThreedimensionalWingStructure2020}.  To determine the aerodynamic performance of a damaged insect wing, flight speed, height, and acceleration of an insect with a naturally damaged wing or with an artificial clipping of a wing were measured \cite{kihlstromWingDamageAffects2021,leroyEffectsNaturalWing2019}. The statistical analysis demonstrated that flight ability diminished when the wing was damaged, and the magnitude varied according to the specific damaged region and the extent of the lost area. With regard to the detailed aerodynamic forces and power consumption of a damaged wing, the measurements and numerical calculations have been conducted on a number of some small insects \cite{mengWingKinematicsMeasurement2023,muijresFliesCompensateUnilateral2017,lyuWingKinematicAerodynamic2020}. 

Among these studies, \cite{lyuWingKinematicAerodynamic2020} explained the reason why a fly's damaged wing can generate a force that is comparable to that of an intact wing through the formation 
of vortex structure at the leading edge. However, there is no existing literature on the flow field of dragonfly damaged wings. Unlike Diptera like fruit fly, dragonflies have two pairs of wings that move independently and a higher Reynolds number, which indicates the existence of a strong forewing-hindwing interaction \cite{hsiehUnsteadyAerodynamicsDragonfly2010,huAerodynamicInteractionForewing2014}. This interaction may change when a wing has lost some area, affecting the other wings force generation even if the other wing is intact. By comparing the detailed flow field and the forces exerted by the wings in damaged conditions, our study identifies the similarities and differences in aerodynamic performance of a dragonfly with damaged wings and intact ones, illustrating the effect of wing area loss on vortex structures in the interacting region and the impact to the undamaged wing therein.

Previous studies on flapping wing vortex and flow field have proposed a variety of mechanisms to explain the force generated by wing stroke and the enhancement of force by wake capture \cite{weis-foghQuickEstimatesFlight1973,saffmanFlowWingAttached1977,
vanveenUnsteadyAerodynamicsInsect2022}. The mechanisms were validated by experiments 
\cite{dickinsonWingRotationAerodynamic1999,srygleyUnconventionalLiftgeneratingMechanisms2002} and 
CFD simulations \cite{gaoInsectNormalHovering2008,kimComputationalInvestigationThreedimensional2011,
engelsBumblebeeFlightHeavy2016,engelsImpactTurbulenceFlying2019,bhatEffectsFlappingmotionProfiles2020}, 
but were rarely combined with damaged wing CFD calculations. Such simulations require a numerical framework that allows easily changing the wing shape; we use a method of the family of IBM (immersed boundary methods) \cite{mittalIMMERSEDBOUNDARYMETHODS2005,schneiderImmersedBoundaryMethods2015} for this task. We combine it with wavelet-based adaptivity that focuses the numerical effort where it is required to ensure a given precision, and automatically adjusts the local (spatial) resolution. 

In the present work we generate a series of damaged dragonfly fore-/hindwing shapes according to damage probability patterns from a field study \cite{rajabiProbabilityWingDamage2017}, and conduct high-resolution CFD simulations for various damaged wings. The aerodynamic forces of damaged wings are calculated to estimate the effect of area loss on flight ability, and we combine it with the analysis of the visualized vortex structures to explain the mechanisms of force decrease observed in damaged cases.
\section{Dragonfly and damaged wing models}\label{sec:Computational_Cfg}
\begin{figure}
\centering
\centerline{\includegraphics[width=1.0\textwidth]{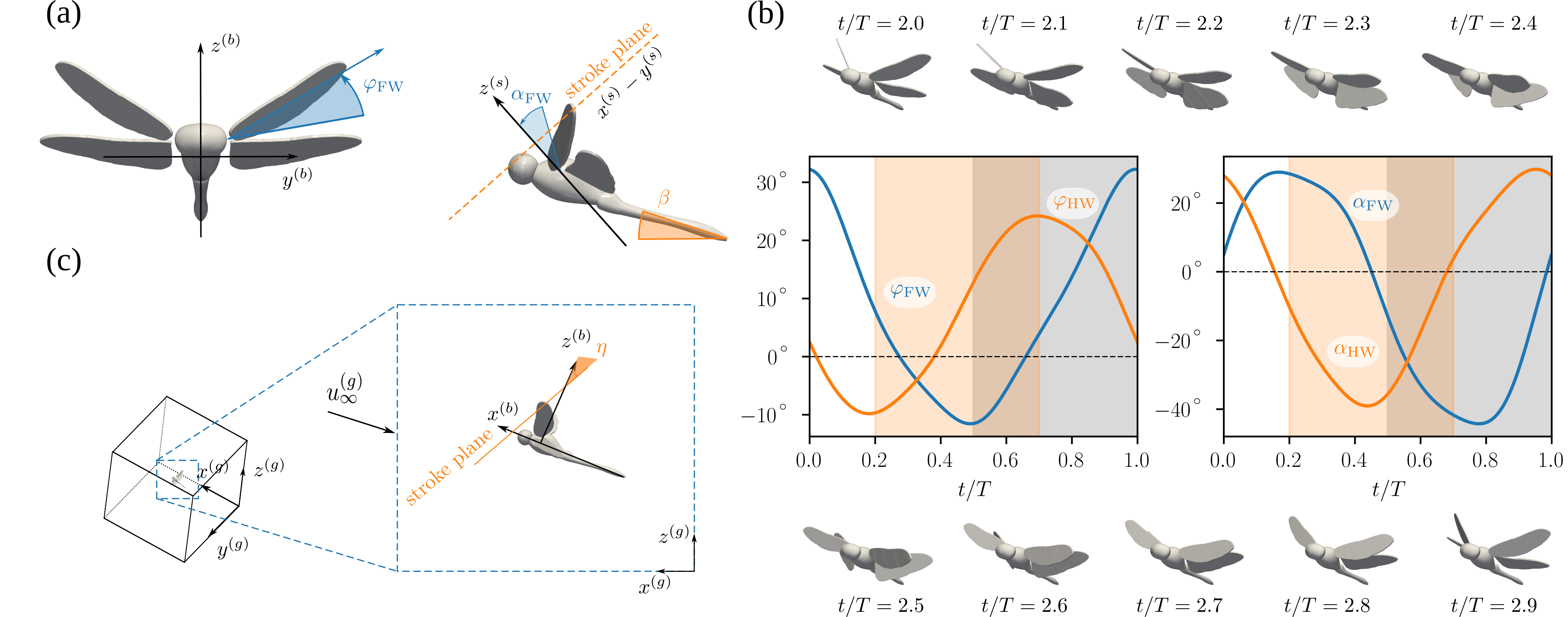}}
\caption{(a) The definition of positional angle ($\varphi$) and feathering angle ($\alpha$). The superscripts $(s)$, $(b)$  and $(g)$ represent the stroke, body and global coordinate system, respectively. The stroke plane is defined as $x^{(s)}-y^{(s)}$ plane and $z^{(s)}$ is perpendicular to it. (b) Rendering of wingbeat cycle and time evolution of kinematics angles. Gray and orange shaded areas indicate the fore- and hindwings upstroke, respectively. (c) Overview of the numerical setup. The dragonfly is at the center of the computing domain; the normalized free stream velocity $u_{\infty }^{(g)}$ is in the $x^{(g)}-z^{(g)}$ plane, with an angle 38$^\circ$ to $x^{(g)}$. The centerline of insect is pitched up by the angle $\beta $ with respect to $x^{(g)}$; the stroke angle $\eta $ is defined with the angle between stroke plane and $z^{(b)}$;}
\label{fig:kinematic_all}
\end{figure}

Our numerical dragonfly model is based on \textit{Pantala flavescens} \cite{heflerAerodynamicPerformanceFreeflying2020}, from which we take the kinematics and geometrical parameters.  We assume flat, rigid wings. Consequently, the wing motion is defined by three kinematic angles: positional angle $\varphi $, feathering angle $\alpha $ and deviation angle $\theta$. However, in forward flight in dragonflies, $\theta $ is generally small and we follow \cite{heflerAerodynamicPerformanceFreeflying2020} and assume that it is negligible. The definitions of the positional and feathering angles are illustrated in figure \ref{fig:kinematic_all}a, and their time evolution is shown in figure \ref{fig:kinematic_all}b, along with a visualization of the wingbeat cycle. For more details and the precise definitions of the different reference frames the reader is referred to \cite{engelsFluSINovelParallel2016}. The Reynolds number, based on the mean wing tip velocity $\overline{u}_{\mathrm{tip},FW}$ and the mean chord length $c_{m,FW}$ (both of the forewing), is $Re=1288$. Assuming flat rigid wings is a simplification of the model presented in \cite{heflerAerodynamicPerformanceFreeflying2020} that included wing twisting. However, the difference in force production is not critical for the present work, as shown by the comparison in Appendix B.

The numerical wind tunnel is constructed as shown in Fig. \ref{fig:kinematic_all}c and we use the same setup for all cases, including damaged ones. In this paper, all length-dependent parameters are normalized by forewing length $R=43.4$ mm, and time-dependent values are normalized by the cycle duration $T=1/f$ , where $f=36.4$ Hz is the flapping frequency. The fluid density $\varrho_\mathrm{air}$ is normalized to unity. We use a $10\times 10\times 10$ large numerical wind tunnel to ensure sufficient space for vortex development. We perform a Galilean change of reference frame and move the air instead of the dragonfly and the free stream velocity vector is $u_\infty^{(g)}=(-0.399,0,-0.310)$, where the superscript $(g)$ denotes a vector in the global coordinate system. The animal is thus in climbing forward flight. The insect's center is at $x_\mathrm{cntr}^{(g)}=(5,5,5)$. Other parameters for the simulations are listed in table \ref{tab:dragonfly_para}. The simulations are carried out with our open-source in-house solver $\mathtt{W} \mathtt{A} \mathtt{B} \mathtt{B} \mathtt{I} \mathtt{T}$, a solver specialized in numerical simulations of flying insects \cite{engelsWaveletAdaptiveMethodMultiscale2021}. We combine wavelet-based adaptivity with the volume penalization method \cite{engelsFluSINovelParallel2016,angotPenalizationMethodTake1999} to enforce the boundary conditions, along with a finite-difference discretization of the governing equations. We solve the Navier-Stokes equations in the artificial compressibility form \cite{engelsWaveletAdaptiveMethodMultiscale2021,ohwadaArtificialCompressibilityMethod2010}.  The wavelet adaptivity dynamically focuses the computational effort where it is required to achieve a prescribed precision. Our simulations are fully resolved, meaning that we do not use turbulence modeling but rather compute all emerging vortices from the Navier--Stokes equations.

\begin{table}
\begin{center}
\def~{\hphantom{0}}
\begin{tabular}{ll}
    Stroke amplitude, forewing, $\varPhi_{FW} $             & $43.70^\circ$\\
    Stroke amplitude, hindwing, $\varPhi_{HW} $& $34.04^\circ$\\
    Wing length of hindwing, $R_{HW}$& 0.924 \\
    Body angle, $\beta$              & $20^\circ$\\
    Stroke angle, $\eta$              & $35.5^\circ$\\
    Normalized kinematic viscosity, $\nu$   & 2.187 $\times 10^{-4}$    \\
    Reynolds number, $Re$                   & 1288   \\
\end{tabular}
\caption{Parameters of dragonfly in forward flight, normalized by forewing length $R$ and cycle duration $T=1/f$.}
\label{tab:dragonfly_para}
\end{center}

\end{table}

\begin{figure}
\centerline{\includegraphics[width=0.65\textwidth]{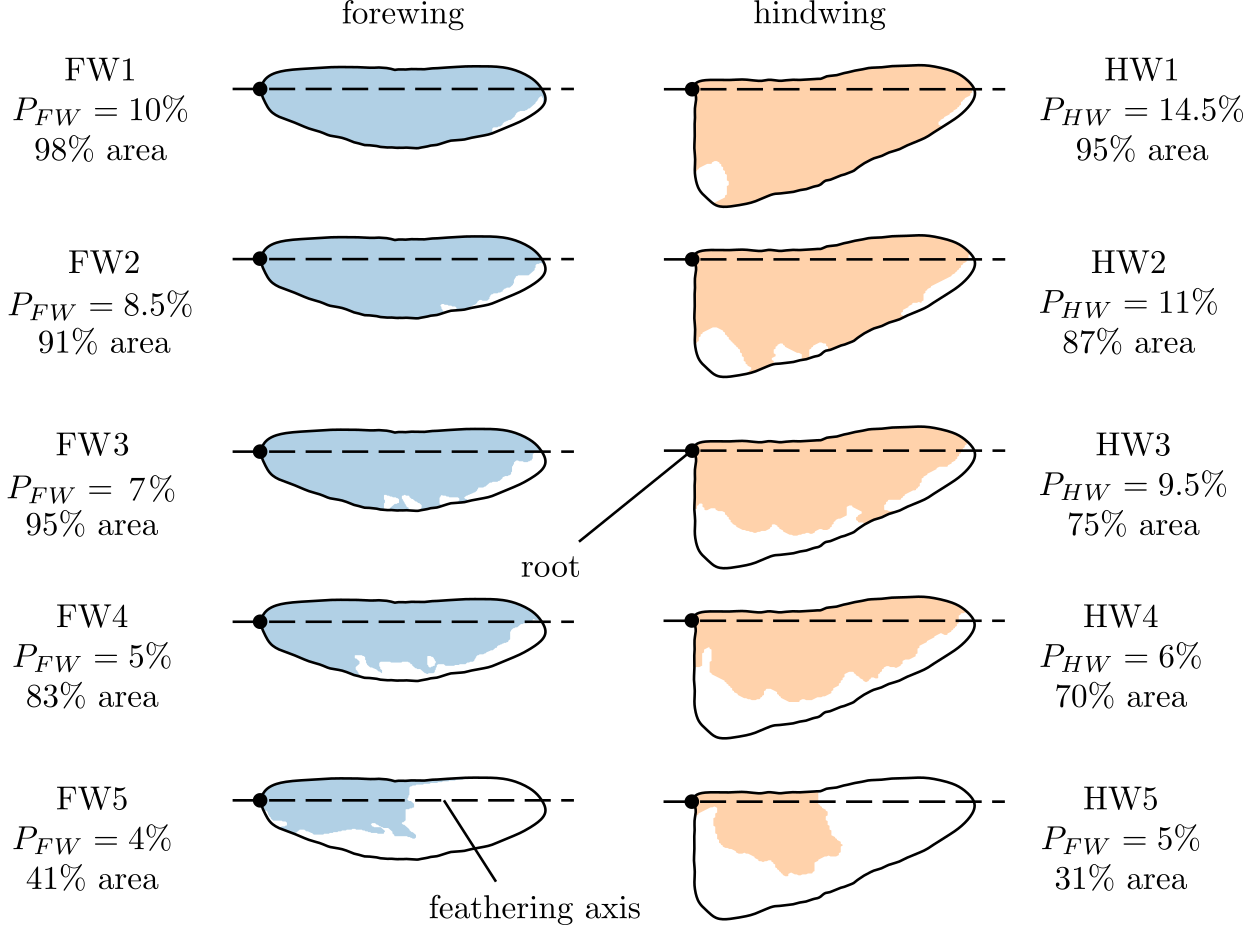}}
\caption{Wing models used in this study, along with the probabilities of damage ($P_{FW}$ and $P_{HW}$) based on the field study of \cite{rajabiProbabilityWingDamage2017}. The various damage patterns show that the damage starts not only from the wing tips, but also near the root on the hindwing.  Animals can sustain a large damage: in the most extreme cases, more than half of the initial area is lost. In the present study, we always combine a damaged fore- or hindwing with an intact hind- or forewing, respectively.}
\label{fig:fig4_DamageCases}
\end{figure}

The shapes of damaged wings are shown in Fig. \ref{fig:fig4_DamageCases}. They are generated based on the probability maps of wing damage presented in \cite{rajabiProbabilityWingDamage2017}. Those data result from a field study in which dragonflies (\emph{Sympetrum vulgatum}) have been caught in the wild and their wing wear has been documented. A probability of $P_{FW}=10\%$ means that 10\% of the animals had lost this part of their wing area. More severe damage was less common, but in the extreme cases, more than half of the wing surface was lost. We pick five probability values for fore- and hindwing. In \cite{rajabiProbabilityWingDamage2017}, a different species has been studied, but no kinematics data are available for that species. However, it has similar, though not exactly identical, wing shapes. The species differ in size: in \emph{P. flavescens}, the wing lengths are 43.4 mm and 40.1 mm (fore- and hindwing), while in \emph{S. vulgatum} they are 28.5 mm and 27.5 mm long. We therefore scale the damage probability maps accordingly before applying it to our model of \emph{P. flavescens}, assuming similar damage patterns in both species.

\section{Results and discussion}\label{sec:results_discussions}

In what follows, we report the results of the damaged cases shown in figure \ref{fig:fig4_DamageCases} and the comparison with the intact case. The kinematics is kept unchanged in all cases to analyze the effect of the damage only on aerodynamic performance. Since the flow field is symmetric in dragonfly flight, without loss of generality, we assume that the damage happens on right fore-/hindwing only. Moreover, the original probability data in \cite{rajabiProbabilityWingDamage2017} do not report differences between the left and right wings. The force and power coefficients are calculated as
\begin{equation}
\begin{aligned}
C_{x,z}^{\mathrm{(FW)}} = F_{x,z}^{\mathrm{(FW)}}/(0.5\rho_{\mathrm{air}}{{u}^{\mathrm{2}}_{\mathrm{tip,FW}}}\,S),\\
C_{P}^{\mathrm{(FW)}}  = P_\mathrm{aero}^{\mathrm{(FW)}}/(0.5\rho_{\mathrm{air}}{{u}^{\mathrm{3}}_{\mathrm{tip,FW}}}\,S)    
\end{aligned}
\label{eq:coeff}
\end{equation}
where ${u}_{\mathrm{tip}}$ is the wing tip velocity of the intact wing; ${S}$ is wing area (damaged or intact); $F_{x,z}$ is horizontal and vertical force, respectively, and $P_\mathrm{aero}$ is the aerodynamic power. Vertical and horizontal forces are along with $z^{(b)}$ and $x^{(b)}$, respectively, and all coefficients are defined in the same way for the hindwing.

\begin{figure}
\centering
\centerline{\includegraphics[width=.85\textwidth]{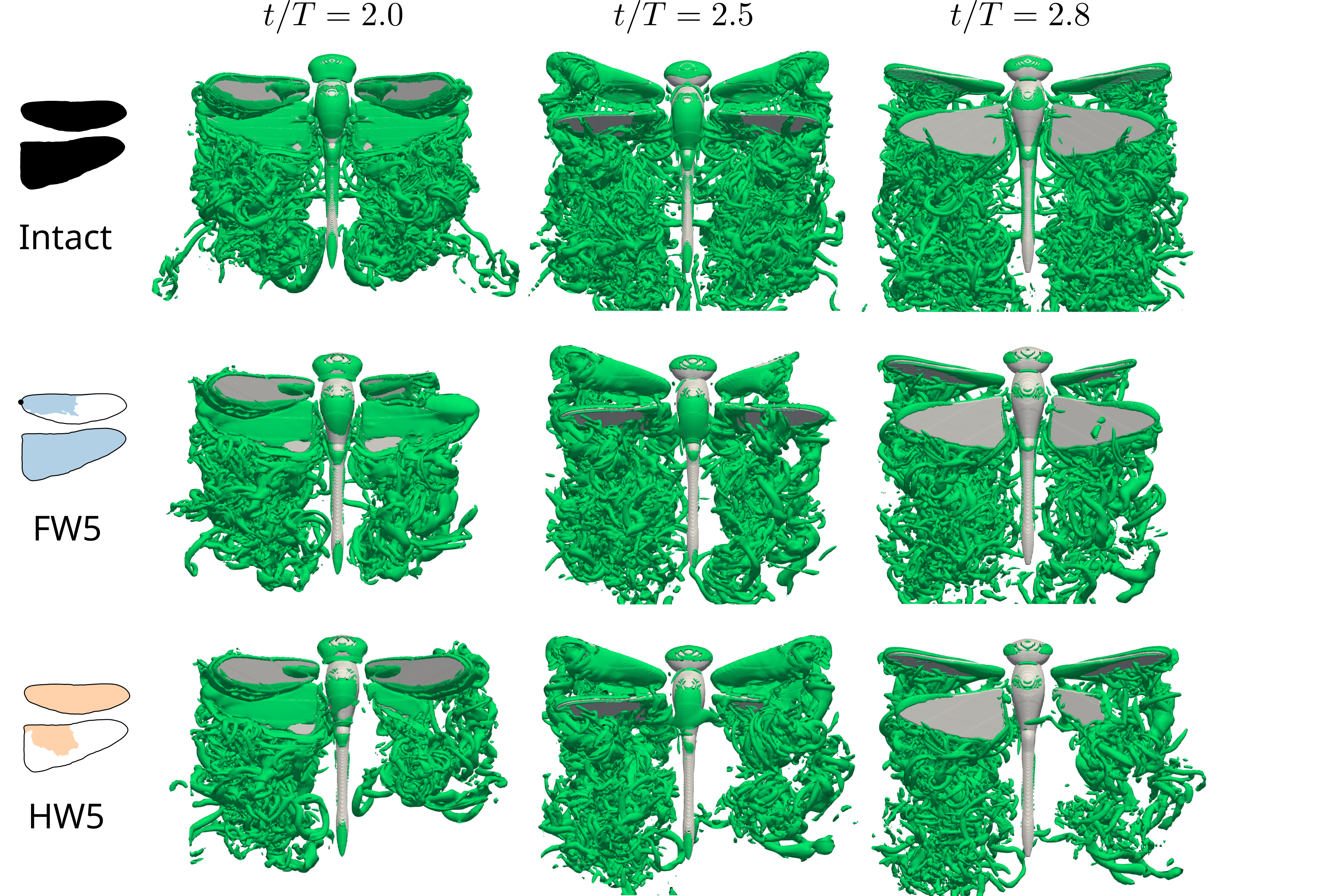}}
\caption{Snapshots of Q-criterion $Q = 50$ contour of three cases. {The dragonfly body models on the left indicate the wing shape case of the row. The vortex figures in the same column are at the $t/T$ on the top.} {The damaged wings are always on the right side of the dragonfly. It appears that the damage only affects the flow on the damaged side, and the other side (left side in the figure) remains the same as in the $\mathsf{Intact}$ case.}}
\label{fig:Qcri-intact-fw4-hw5}
\end{figure}

We use Q-criterion contours \cite{huntEddiesStreamsConvergence1988} to visualize the vortex structures in the flow field around the wings. Figure \ref{fig:Qcri-intact-fw4-hw5} shows a comparison between the intact case, the case with the most damaged forewing ($\mathsf{FW5}$), and the case with the most damaged hindwing ($\mathsf{HW5}$), through a series of consecutive snapshots. The wake on the damaged side (the right side) clearly has a different structure compared to the intact wing case and also varies considerably depending on whether the damaged wing is the forewing or the hindwing. In the three snapshots of the $\mathsf{FW5}$ case, the tip vortex (TV) of the forewing is broken near the damaged region, consequently shifting its interaction with the hindwing towards the root. This may result in a force difference of the hindwing on the same side of the damaged forewing even it the hindwing itself is intact. In contrast, in the $\mathsf{HW5}$ case, the TV shed by the forewing goes past the hindwing without obstacle, due to the large area loss on the hindwing. The shed hindwing vortex becomes less developed and does not interact with the dragonfly body near the abdomen. We note that in all the snapshots of these two largely damaged cases, the vortex of the undamaged side remains the same as that of the intact wing case, which means that the loss of wing area only affects the flow field on the side of the damaged wing (the right wing here). Hence, all the discussions in the rest of the paper concentrate on the right side (the damaged side).

\subsection{Effect of the wing damage pattern}\label{sec:effect_pattern}

\subsubsection{Forewing damage}

\begin{figure}
\centerline{\includegraphics[width=.75\textwidth]{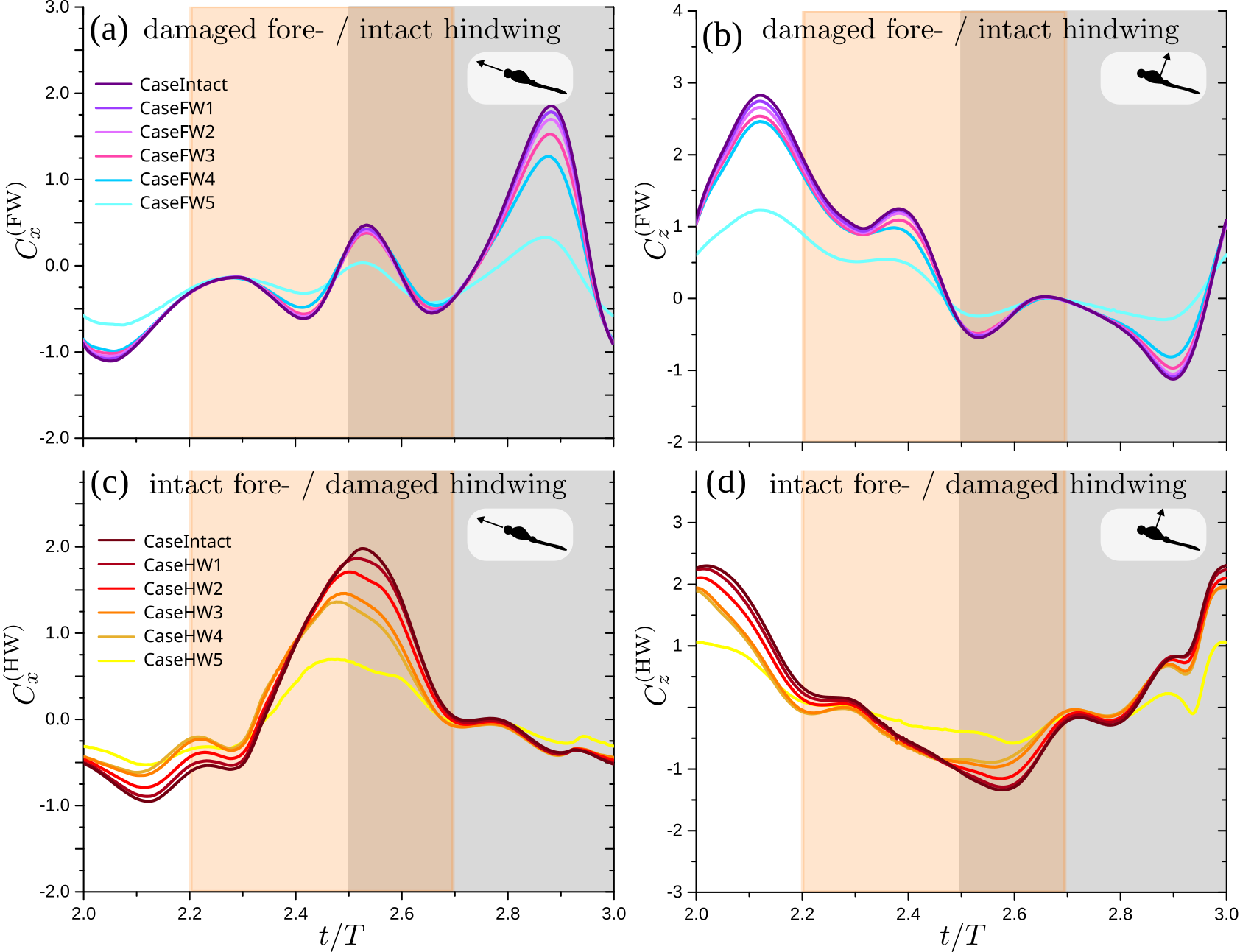}}
\caption{Instantaneous aerodynamic force coefficients. Data from simulations with damaged forewing (top row) and damaged hindwing (bottom row). Shown are horizontal (left column) and vertical (right column) force coefficients (see insets) of fore- (top) and hindwing (bottom). Grey (orange) areas indicate forewing (hindwing) upstroke.}
\label{fig:forces_damage_pattern}
\end{figure}

{We start with the discussion of the aerodynamic performance in damaged forewing cases{, where the hindwing is intact.}} Figures \ref{fig:forces_damage_pattern}a, \ref{fig:forces_damage_pattern}b show the variation of the horizontal and vertical force coefficients during one flapping period for all damaged forewing cases as well as for the intact case. 

It is obvious that the force coefficients decrease as the damaged zone becomes larger. In the extreme case $\mathsf{FW5}$, where only a small region close to the wing root remains, the magnitude of the horizontal and vertical forces of the forewing drop to very low values. The drop is especially pronounced in the parts of the cycle that produce the largest force peaks in the intact and less damaged wings: the largest drop of the horizontal force coefficient occurs during the upstroke, when $t/T \approx 2.9$, while that of the vertical force coefficient occurs during the downstroke, when $t/T \approx 2.1$. 

\begin{figure}
\centerline{\includegraphics[width=0.7\textwidth]{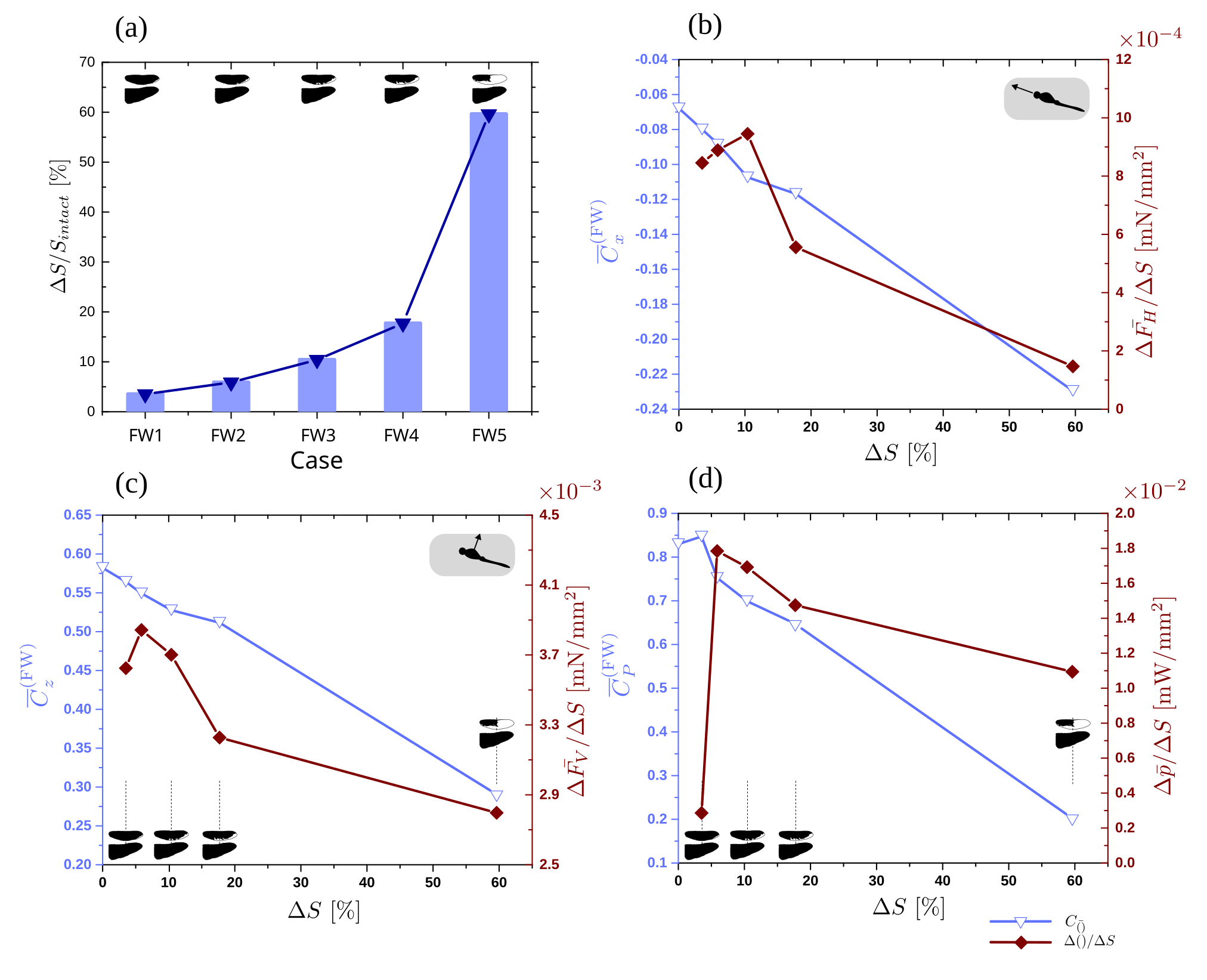}}
\caption{{\textbf{Average aerodynamic coefficients  for the damaged fore-/intact hindwing cases}: (a) $\Delta S$ is the wing area loss and $\Delta S/S_{intact}$ is the ratio between the area loss and the surface of the intact wing. $\Delta$ represents the difference between the intact case and the damaged cases. (b) Horizontal force, (c) vertical force, and (d) aerodynamic power  coefficients, as a function of the area loss. In figures (b-d), in addition to the average coefficients shown in blue (scale on the vertical axis on the left of each plot), the ratio of the force $\Bar{F}$ (or power $\Bar{P}_\mathrm{aero}$) to the area loss of each damaged forewing (scale on the vertical axis on the right of each plot). }
}
\label{fig:average_p_F_fw}
\end{figure}

In order to assess the global effect of the wing damage, we now examine the average values of the aerodynamic force and power coefficients.  Figure \ref{fig:average_p_F_fw} shows the average horizontal and vertical forces and the power coefficients for all cases as a function of the lost wing area $\Delta S$ of each damaged forewing. 
Examining again the case with the largest damage ($\mathsf{FW5}$), it is not surprising that the average force/power coefficients are very small compared to the intact wing and other damaged cases, as the remaining wing area is only 40\%. 
However, it should be noted that the average horizontal force 
is actually slightly higher ($\approx3\%$) in this case than in $\mathsf{FW4}$, which has less damage. This apparently counterintuitive result can be explained recalling that in the forward flight of this study the thrust force is the result of the aerodynamic forces on both pairs of wings. The forewings generate on average a negative horizontal force in one cycle (i.e., a drag force), while the contribution to the horizontal thrust is mainly from the hindwings (see figure \ref{fig:average_p_F_hw}). Compared to $\mathsf{FW4}$ and $\mathsf{FW3}$, the form drag is smaller in $\mathsf{FW5}$ because a large proportion of the wing is removed, resulting in a smaller drag in the horizontal direction. 

The ratios $\Delta \overline{F} / \Delta S$ and $\Delta \overline{P}_{_\mathrm{aero}}/ \Delta S$ are calculated to investigate the impact of the damaged region on the flight performance of the wing (figures \ref{fig:average_p_F_fw}b-d, {right y-axis}). The ratios decrease in the two middle cases ($\mathsf{FW3}$ and $\mathsf{FW4}$) in the figure, indicating that, although the area loss increases, the impact per unit area is reduced. Referring to Figure \ref{fig:fig4_DamageCases}, the lost area is concentrated on the trailing edge in $\mathsf{FW4}$ and $\mathsf{FW3}$, which appears to be not that vital for the dragonfly forewing. 
But for the region near the wing tip, the situation is different. When comparing the cases $\mathsf{FW1}$ and $\mathsf{FW2}$, there is an increase (a sharp increase for power) between the two cases. The area loss regions are close to the wing tip, which strongly influences the tip vortex, resulting in a stronger sensitivity of power to area loss.


\begin{figure}
\centering
\centerline{\includegraphics[width=.6\textwidth]{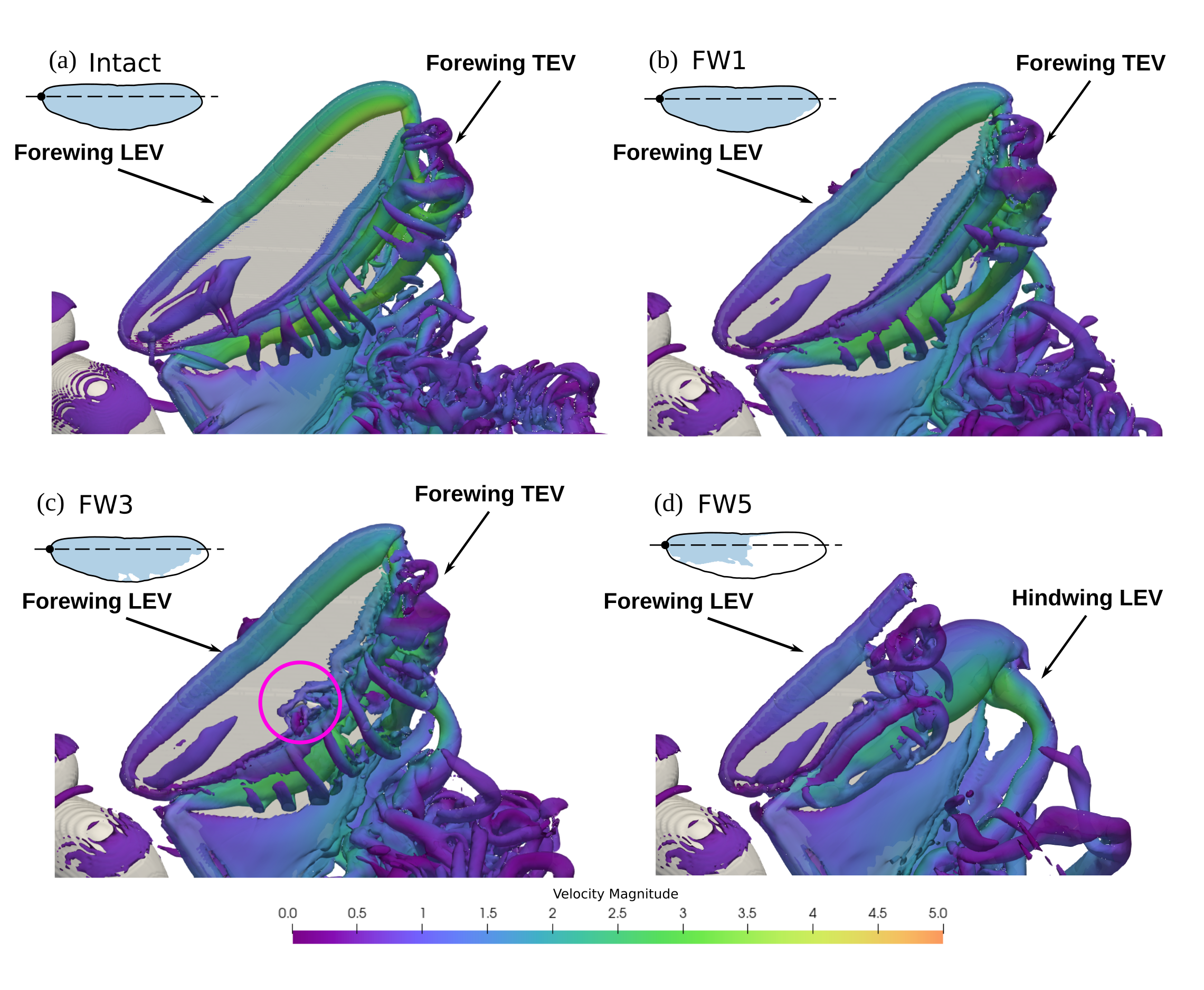}}
\caption{{Simulations with damaged fore- and intact hindwing, } vortex structure around right forewing of intact and  three typical damaged cases at $t/T=2.1$. The wake structure is visualized with Q-criterion $Q = 50$ and colored with velocity. The top left insect in each panel indicates the corresponding wing shape. There is a hole in the pink circle and we can see the small deterioration of the vortex around the hole.}
\label{fig:fig_LEV_intact-fw4-10}
\end{figure}

In flapping-wing aerodynamics, the leading edge vortex (LEV) is associated with the mechanism of force generation through the production of a low pressure region close to the wing during a part of the flapping cycle \cite{vandenbergVortexWakeHovering1997,dickinsonWingRotationAerodynamic1999,
birchForceProductionFlow2004,bode-okeFlyingReverseKinematics2018,vanveenUnsteadyAerodynamicsInsect2022}.  Figure \ref{fig:fig_LEV_intact-fw4-10} illustrates the LEV and trailing edge vortex (TEV) of the intact case and the $\mathsf{FW1}$, $\mathsf{FW3}$, and $\mathsf{FW5}$ cases with damaged forewings. The snapshots correspond to the moment of maximum vertical force production (the largest force peak in figure \ref{fig:forces_damage_pattern}b at $t/T=2.1$). Compared to the $\mathsf{Intact}$ case, the LEV of the $\mathsf{FW5}$ case remains complete from root to mid-wing, but disappears in the region close to the wing tip due to the large damage. For the other damaged forewing cases, the leading edge is not damaged and thus enables the presence of an LEV as strong as in the intact wing, which supports the basic force generation ability. This can explain the relatively small change in the force coefficients in figures \ref{fig:forces_damage_pattern}a and \ref{fig:forces_damage_pattern}b for the $\mathsf{FW1}$ and $\mathsf{FW3}$ cases with respect to the intact case. On the other hand, the vortex distortion occurs when the trailing edge is damaged, especially when the shape of the damaged wing is irregular (pink circle in Figure \ref{fig:fig_LEV_intact-fw4-10}). The distortion of the trailing edge vortex (TEV) represents another potential reason for the decrease in force in damaged wings.

\subsubsection{Hindwing damage}

\begin{figure}
\centerline{\includegraphics[width=0.7\textwidth]{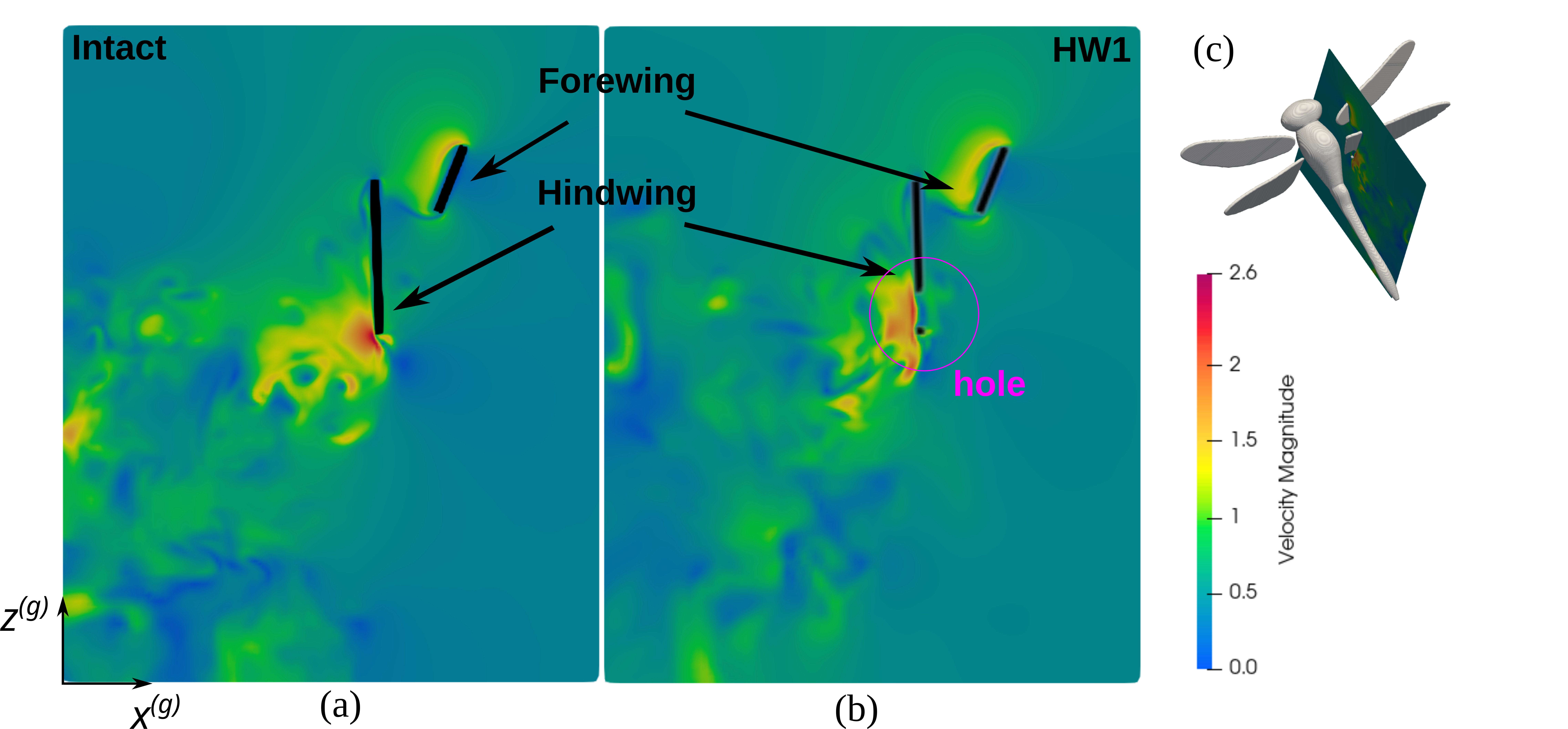}}
\caption{{Simulations with intact fore- and damaged hindwing: } velocity magnitude at $t/T=2.5$. (a) $\mathsf{Intact}$; 
(b) $\mathsf{HW1}$; (c) The location of the flow field slice, which goes cross the hole at hindwing root and is perpendicular to $y^{(g)}$ axis. The deterioration of the velocity field around the hole and the perturbation of the hole to downwash flow development can be seen.}
\label{fig:slice-hw145-intact}
\end{figure}
\begin{figure}
\centerline{\includegraphics[width=0.7\textwidth]{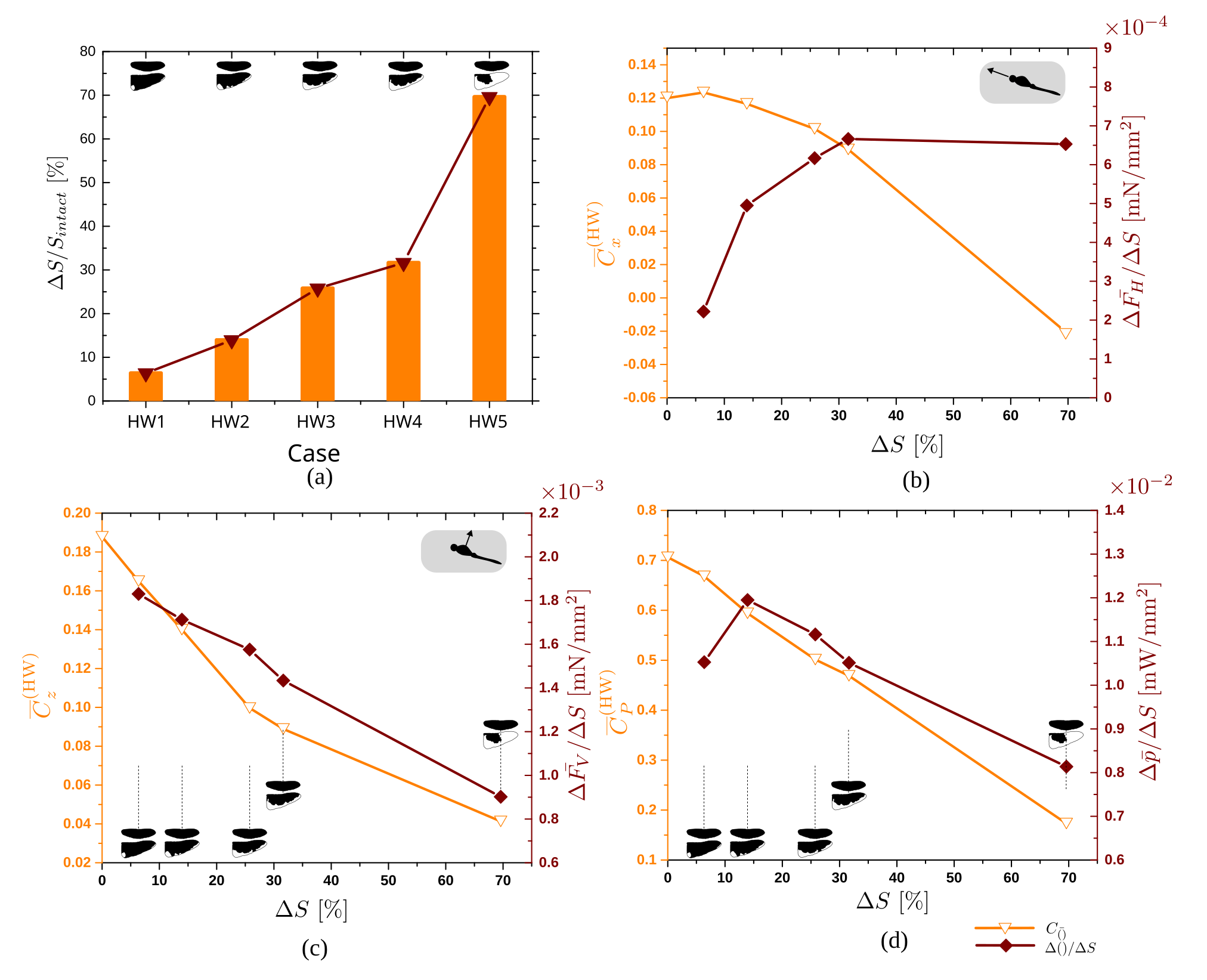}}
\caption{\textbf{Average aerodynamic coefficients  for the damaged hindwing cases}: (a) Wing loss areas for each case. (b) Horizontal force, (c) vertical force, and (d) aerodynamic power  coefficients, as a function of the area loss. The notations are analog to those of figure \ref{fig:average_p_F_fw}}.
\label{fig:average_p_F_hw}
\end{figure}

Both the intact wing shape and the general damage patterns for the hindwing are very different from those of the forewing, and this is reflected in the time evolution of the force coefficients (Figs. 
\ref{fig:forces_damage_pattern}c and \ref{fig:forces_damage_pattern}d). The horizontal force coefficient of $\mathsf{HW1}$ has a drop during the upstroke, at $t/T \approx 2.5$. For the damaged forewing, the slightly damaged case $\mathsf{FW1}$ also has a decrease in the horizontal force coefficient during the upstroke, but the difference is much smaller. The damaged pattern of $\mathsf{HW1}$ is similar to that of $\mathsf{FW1}$ at the wing tip, but $\mathsf{HW1}$ has a large hole on the trailing edge at the wing root, and the damage expands from the hole in the following cases with more damage. Figure \ref{fig:slice-hw145-intact} shows the perturbation in the flow field produced by this perforation. Compared with the $\mathsf{Intact}$ case, the existence of a hole in the wing root makes the velocity field more disordered. The high-velocity flow near the root of the trailing edge cannot be sustained and dissipates around the hole. The resulting downwash flow is therefore weaker than in the $\mathsf{Intact}$ case.

When the hindwing damage becomes more severe, the peaks in force coefficients gradually decrease in magnitude: for the horizontal force coefficient, this happens during the upstroke, while for the vertical one it happens during the downstroke. In case $\mathsf{HW3}$, the force coefficient peaks have a strong decrease compared to the intact hindwing. The area loss is less than 1/3, but focused around the trailing edge. 
As in the forewing case $\mathsf{FW2}$, the trailing edge remains relatively functional, leading to an aerodynamic force approximately equal to that of the intact wing. For the most damaged wing (case $\mathsf{HW5}$) more than half of the wing area is missing and, as in the case of the largely damaged forewing $\mathsf{FW5}$, the wing can hardly generate forces in either direction.

Analogously to Figure \ref{fig:average_p_F_fw} for the damaged forewing cases, Figure \ref{fig:average_p_F_hw} shows the average horizontal and vertical forces and the power coefficients as a function of the lost wing area $\Delta S$ for the damaged hindwing cases, where $\Delta S=0$ corresponds to the intact wing. When the hindwing is largely damaged ($\mathsf{HW5}$ case), the average horizontal force coefficient $\Bar{c}_{F_{H}}$ can become negative (Figure \ref{fig:average_p_F_hw}b) and the average vertical force coefficient $\Bar{c}_{F_{V}}$ is also very small (Figure \ref{fig:average_p_F_hw}c). In this case, the dragonfly in nature needs to adjust the kinematics, such as increasing the flapping frequency or flapping amplitude, to sustain itself in air against gravity. 

For the case $\mathsf{HW1}$ (the leftmost case in figures \ref{fig:average_p_F_hw}c-d, where the wing has only an area loss of $\approx 6.4$\%), the horizontal force remains the same as for the intact wing, but the vertical force drops significantly, as does the average power, although less pronounced. The center of the damaged region in this case is on the trailing edge near the wing root, which implies that the hole in the hindwing and the corresponding perturbation to the flow (figure \ref{fig:slice-hw145-intact}) considerably affect the vertical force and power, but have a negligible impact on the horizontal force ($\Bar{c}_{F_{H}}$ is even slightly larger than in the $\mathsf{Intact}$ case).
Comparing the two cases $\mathsf{HW1}$ and $\mathsf{HW2}$, the differences per area of $\Bar{F}_{H}$ and $\Bar{P}_\mathrm{aero}$ change significantly, especially for $\Delta \Bar{F}_{H}/\Delta S  $. Figure~\ref{fig:fig4_DamageCases} shows for these cases that the damage expands in two regions: the wing tip and trailing edge, which can both be responsible for the diminishing  $\Bar{F}_{H}$ and $\Bar{P}_\mathrm{aero}$. Now, concerning cases $\mathsf{HW3}$ and $\mathsf{HW4}$, where the wing shapes differ mostly at the trailing edge, the values of $\Delta \Bar{F}_{H}/\Delta S$ and $\Delta \Bar{P}_{_\mathrm{aero}}/\Delta S$ are close to each other, which implies that the horizontal force and power are more sensitive to damage at the wing tip than at the trailing edge, similar to the observation of the damaged forewing cases. 

\begin{figure}
\centerline{\includegraphics[width=0.95\textwidth]{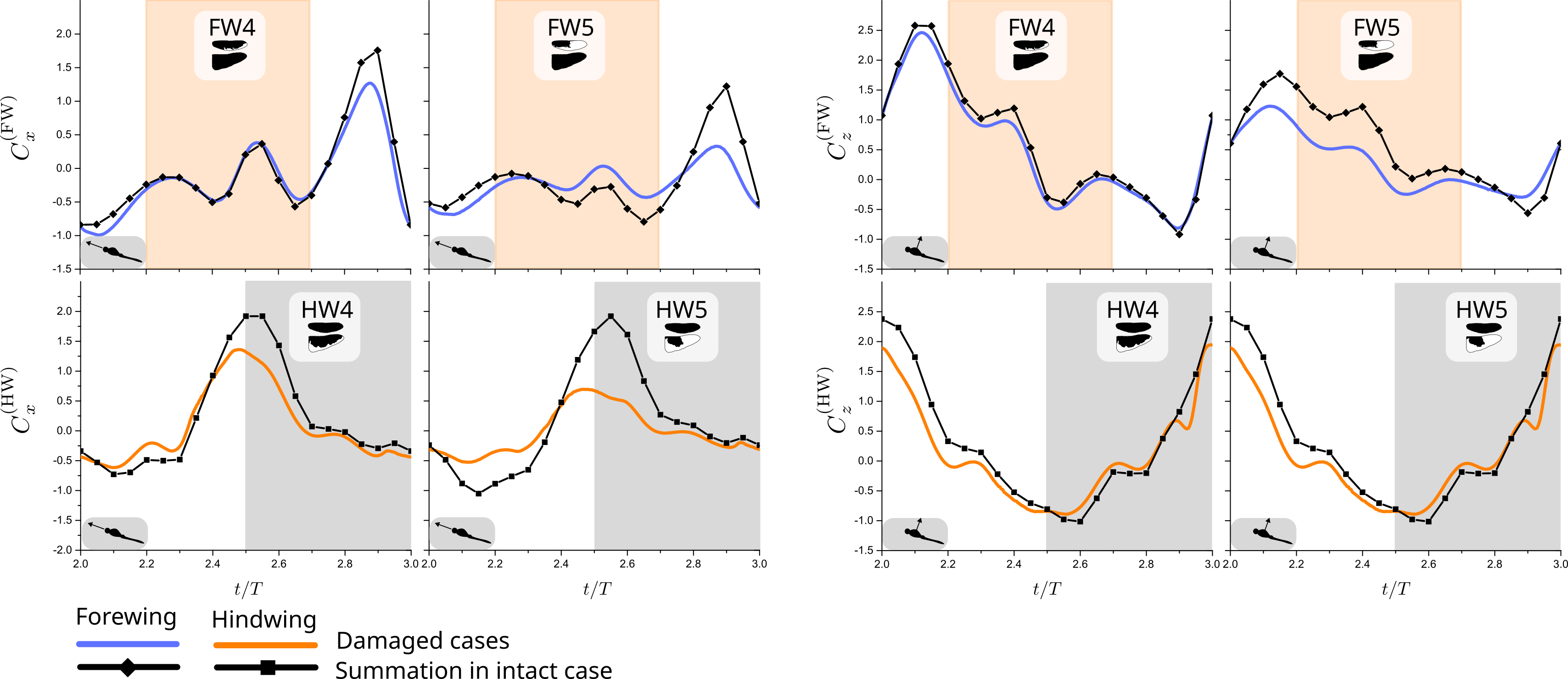}}
\caption{Horizontal and vertical force coefficients for cases $\mathsf{FW4}$, $\mathsf{HW4}$, $\mathsf{FW5}$, and $\mathsf{HW5}$ (colored lines) vs. an hypothetical force coefficient computed from the intact wings flow field but where the force integration is performed only over the area that corresponds to each damaged case (black lines, see text). Grey (orange) areas indicate forewing (hindwing) upstroke. Additional comparisons of all damaged cases can be found in the supplementary materials.}
\label{fig:damagedvsintact}
\end{figure}

The reduction in force in damaged wings compared to an intact wing results firstly from the loss of wing area and secondly from differences in the flow field; both factors are coupled. To disentangle their respective roles, we compare in Figure~\ref{fig:damagedvsintact} the force coefficients of the damaged wing in each case with an hypothetical coefficient computed from the flow field from the intact wing simulation, but where the force integration is performed only on the part of the wing that corresponds to the damaged shape. In practice, we take 21 snapshots in one cycle and sum the forces of the finite blocks obtained in the intact simulation.
Using this strategy, we can remove the effect caused by area loss and just determine the force reduction produced by changes in the flow field. The comparisons of four selected damaged cases are shown in Figure~\ref{fig:damagedvsintact}. In the less damaged cases ($\mathsf{FW4}$, $\mathsf{HW4}$), the force coefficients are close to the hypothetical coefficients computed from the limited integration of the intact wing forces---apart from a smaller peak of the hindwing horizontal force coefficient at the upstroke to downstroke transition time. For the two severely damaged cases $\mathsf{FW5}$ and $\mathsf{HW5}$, the change in the flow field not only reduces the force generation but also decreases the fluctuation amplitude. 

\begin{figure}
\centerline{\includegraphics[width=.75\textwidth]{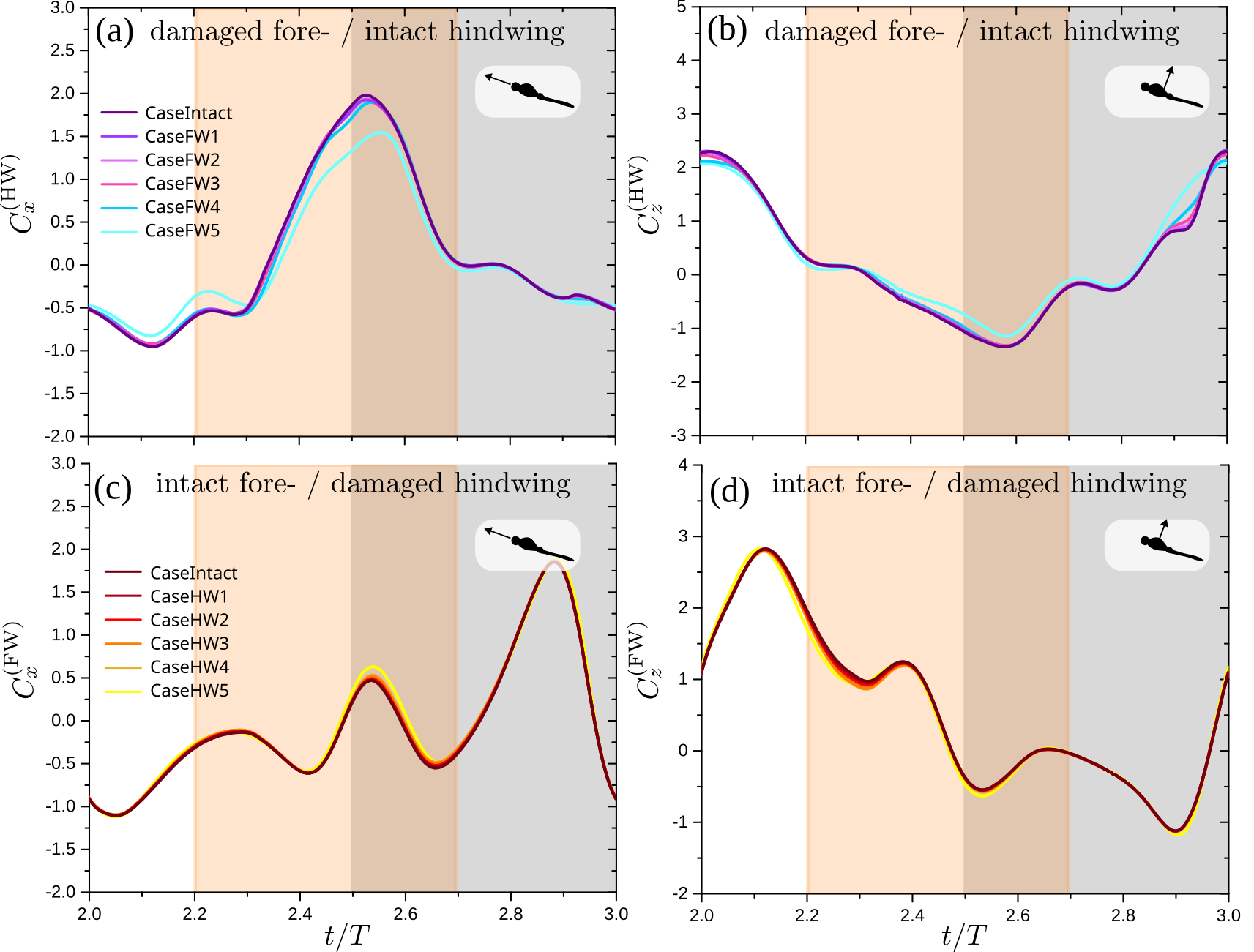}}
\caption{Ipsilateral fore-/hindwing aerodynamic force coefficients in different damaged cases. (a-b) are force coefficients of the intact hindwing which is on the same side of the damaged forewing, while (c-d) are for the intact forewing on the same side as the damaged hindwing. Grey (orange) areas indicate forewing (hindwing) upstroke.}
\label{fig:force-wing-wing-interaction}
\end{figure}

\subsection{Effect of wing-wing interactions}\label{sec:effect_fhi}
Up to now we have discussed forces on each wing separately, ignoring the obvious fact that the dragonfly has two pairs of wings, which means that fore- and hindwing interact. The hindwing is in the downstream flow of the forewing; therefore, changes in the flow field produced by the forewing and the different vortex structures generated by the different damaged wing shapes will affect the flow environment of the hindwing, resulting in a force difference compared to the ipsilateral hindwing of the $\mathsf{Intact}$ case. 

\begin{figure}
\centerline{\includegraphics[width=.85\textwidth]{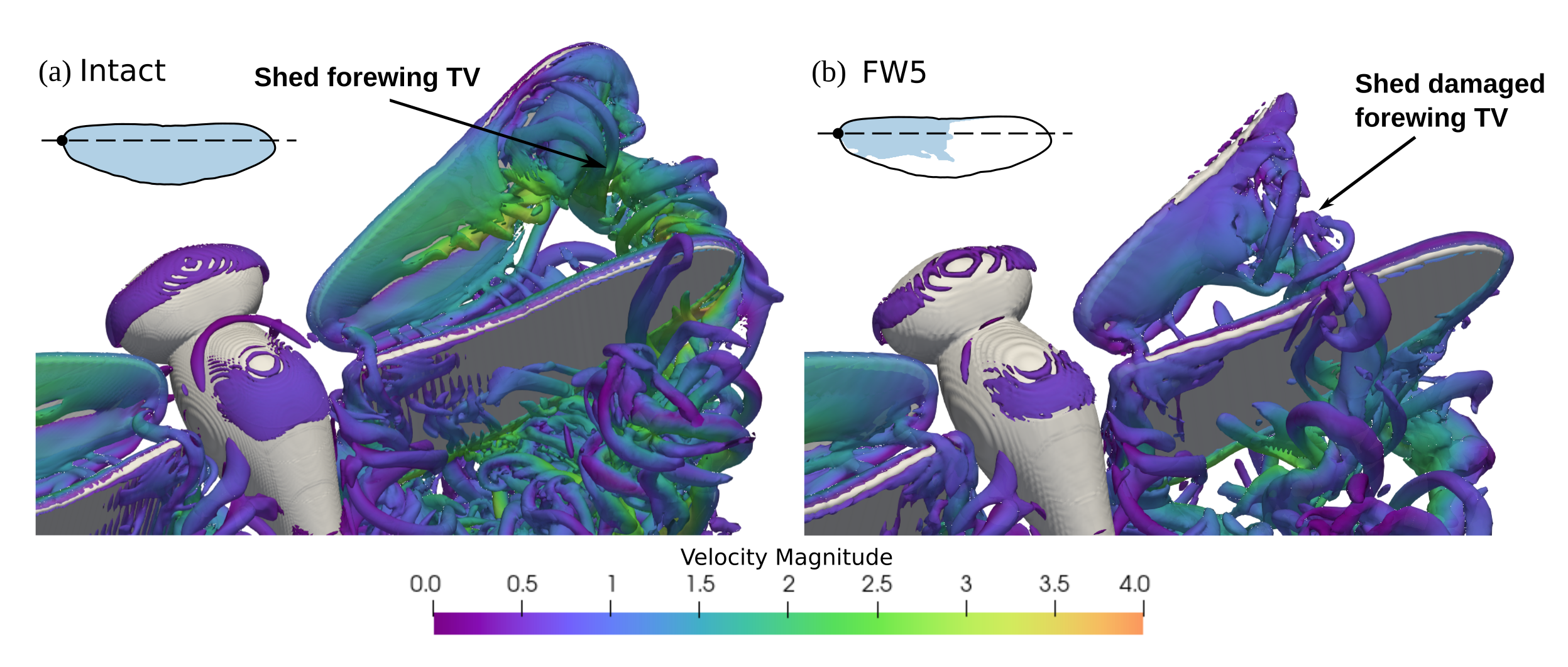}}
\caption{Forewing-hindwing interaction of intact and largely damaged forewing case $\mathsf{FW5}$ at $t/T=2.5$. The wake structure is visualized with Q-criterion $Q = 50$ and with velocity contour. The black wing at top left in each figure indicates the corresponding wing shape. The videos of the interaction can be found in supplementary materials.}
\label{fig:fig_FHI_intact-fw4}
\end{figure}

Figures \ref{fig:force-wing-wing-interaction}(a-b) illustrate the horizontal and vertical force coefficients of the ipsilateral hindwing to the damaged forewing in all damaged cases and compare them with the $\mathsf{Intact}$ case. Although the increase in lost area on the forewing results in a reduction in the aerodynamic forces of the forewing, a notable change of the ipsilateral hindwing force coefficient is only observed in the largely damaged forewing case $\mathsf{FW5}$, especially for the horizontal force. 
Wake capture has been described in the literature as an unsteady mechanism of aerodynamic force production of flapping wings \cite{dickinsonWingRotationAerodynamic1999,birchSpanwiseFlowAttachment2001, srygleyUnconventionalLiftgeneratingMechanisms2002,vanveenUnsteadyAerodynamicsInsect2022}. This mechanism is typically referred to when the wake generated by the previous stroke can provide gain to the force generation of the flapping wing. For 
Odonata with their two pairs of wings, wake capture happens between the wake of the forewing influences the force of the hindwing. Figure \ref{fig:fig_FHI_intact-fw4} presents a comparison of the wake structure between a forewing and the ipsilateral hindwing  for the $\mathsf{Intact}$ case and $\mathsf{FW5}$ damaged case at $t/T=2.5$ (see videos in the supplementary materials). In the $\mathsf{Intact}$ case in figure \ref{fig:fig_FHI_intact-fw4}a, the shed forewing TV goes through the wing tip of the hindwing, which is the main region for force generation \cite{engelsThreedimensionalWingStructure2020}. While in figure \ref{fig:fig_FHI_intact-fw4}b for the damaged case, since the forewing wing tip is lost, the shed vortex of the new forewing tip is weaker, and it interacts with the hindwing in its middle region. The reduction of forewing-hindwing interaction and the displacement of the interacting region results in the decrease of force of the ipsilateral hindwing when the forewing is largely damaged. As with the other cases, the change of the forewing vortex structure is confined to the region close to the trailing edge of the forewing and the vortex acting on the ipsilateral hindwing remains constant, which does not lead to a force decrease of the hindwing.

\begin{figure}
\centerline{\includegraphics[width=0.85\textwidth]{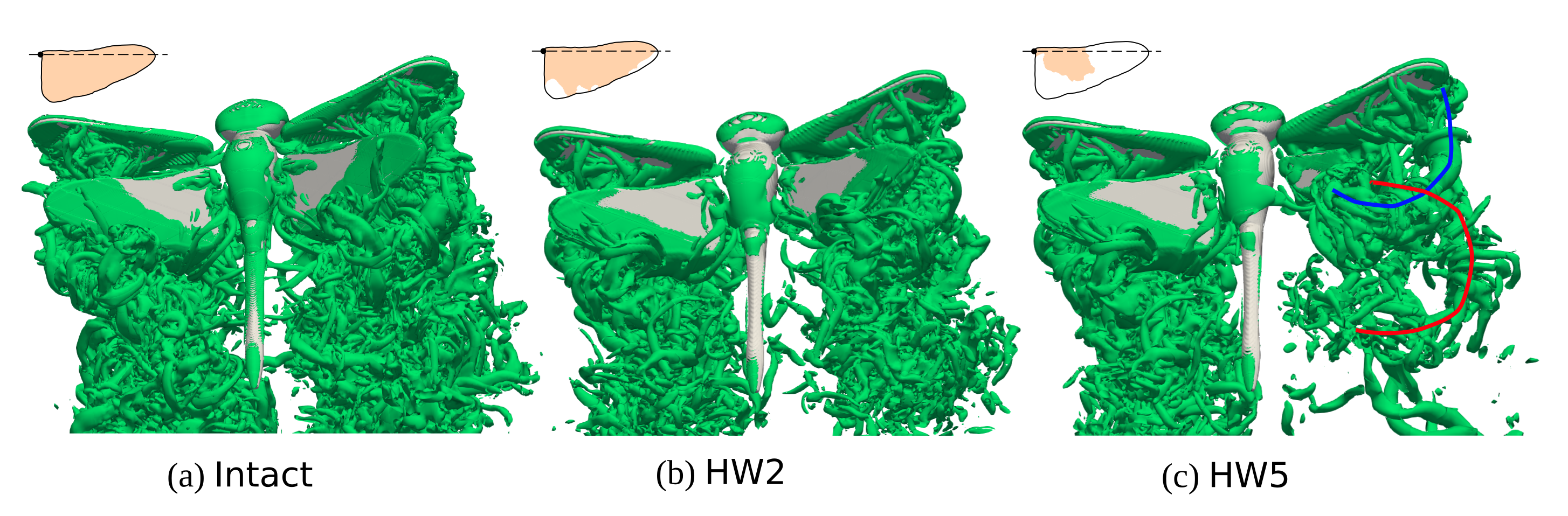}}
\caption{Snapshots of Q-criterion $Q = 50$ contour of intact and damaged hindwing cases. The development trajectory of these vortices in $\mathsf{HW5}$ are drawn with blue and red line, respectively. All snapshots are at $t/T=2.7$, which is the end of hindwing upstroke. {The top left black wing indicates the corresponding damaged wing shape.}}
\label{fig:TV-hind-t2_7-intact-damaged}
\end{figure}

Contrary to the ipsilateral hindwing of a damaged forewing, the ipsilateral forewing in a damaged hindwing case flaps in the upstream of the hindwing. This indicates that the area loss of the hindwing should have less effect on the forewing. Figure 
\ref{fig:force-wing-wing-interaction}(c-d) shows that all the ipsilateral forewings in various damaged hindwing cases generate similar forces as in the $\mathsf{Intact}$ case. From the Q-criterion contour in figure \ref{fig:TV-hind-t2_7-intact-damaged}, it can be seen that the forewing does not interact with 
the hindwing shed vortex. The shed hindwing TV is always beneath the forewing, as in $\mathsf{HW5}$ 
(figure \ref{fig:TV-hind-t2_7-intact-damaged}c). In this case, since the hindwing is largely damaged, the development trajectory of the shed hindwing TV 
can be clearly observed (red line in figure \ref{fig:TV-hind-t2_7-intact-damaged}c).


\section{Conclusions}\label{sec:conlusions}
A series of CFD simulations have been conducted on a model dragonfly to examine the aerodynamic effect of wing damage. Different degrees of forewing and hindwing damage were examined using typical damage patterns observed in nature to determine the modification of aerodynamics forces and flow fields with respect to a reference case with intact wings. The methodological approach of the paper was to maintain the same wing beat kinematics for all simulations, in order to quantify the impact of the damaged wing shapes with various intensities of damage on the capacity to generate aerodynamic forces, and to elucidate the mechanisms behind the force reduction.

Although, as expected, the expansion of damage resulted in a decrease of forces due to the reduction of wing area, the difference of force coefficient decrease cannot be simply explained by the percentage of wing area loss. The pattern of damage plays an important role. The results show that, apart from the largely damaged cases $\mathsf{FW5}$ and $\mathsf{HW5}$, in all other cases where the leading edge was kept undamaged, a strong LEV continued to be present, ensuring the basic ability of damaged fore- and hindwings to produce forces. 

From the comparison of average forces and power coefficient in the damaged cases, we found different sensitivities of aerodynamic performance depending on the region where the wing damage is located. For both the forewing and the hindwing, the force generation and power consumption were both very sensitive to the damage at the wing tip. 
Increasing damage at the wing tip induces a severe drop in flight performance, especially seen in the aerodynamic power. Hindwing damage at the root of the trailing edge, which had the highest probability in the damaged patterns from the literature, can largely impact the vertical force and power coefficient even though it only represents less than 10\% of the whole wing area in some damaged cases. While this type of damage diminishes the vertical force and could challenge the flight of a dragonfly against gravity, the horizontal force was almost not affected. This directional inhomogeneity of force generation means that in practice a dragonfly could compensate for wing damage not only by changing the wing beat kinematics, but also by reorienting its body and stroke plane with respect to the vertical.

Due to the existence of forewing-hindwing interaction, the global impact of damage on one wing can affect the force production of the other wings. We found that 
the ipsilateral hindwing of a largely damaged forewing generated less horizontal force in the upstroke due to the reduction of forewing-hindwing interaction. In $\mathsf{FW5}$ where the forewing lost more than half of the wing, the shedding of the forewing TV shifted, resulting in a change of the interacting location on the ipsilateral hindwing and the wake captured by the hindwing. This shift weakened the enhancement of the hindwing force from the shed forewing vortex, leading to a force decrease of the hindwing.

The present study of damaged dragonfly wings contributes to a better understanding of forewing-hindwing interactions and to better force predictions for insect inspired flapping-wing micro air vehicles with unexpected damage.

\section*{Appendix A. Numerical method}\label{appendix-A}
To conduct the simulations in this study, we use our previously published CFD code 
$\mathtt{W} \mathtt{A} \mathtt{B} \mathtt{B} \mathtt{I} \mathtt{T}$ (https://github.com/adaptive-cfd/WABBIT) for calculating 
the flow field and aerodynamic forces. A brief introduction of the methods used in 
$\mathtt{W} \mathtt{A} \mathtt{B} \mathtt{B} \mathtt{I} \mathtt{T}$ is described in this section. 
For more details, the reader is referred to \cite{engelsFluSINovelParallel2016,engelsWaveletAdaptiveMethodMultiscale2021}.

$\mathtt{W} \mathtt{A} \mathtt{B} \mathtt{B} \mathtt{I} \mathtt{T}$ uses three-dimensional Cartesian grid to deal with the complex geometry of the dragonfly models, especially for the irregular damaged wing shapes. 
The solid surface is extracted from orthogonal grid with refinement rather than generating bodyfitted grid using unstructured grid, which enables the code to contribute an accurate flow field for irregular damaged wing with high flexibility. The solid domains are regarded as porous media in the Cartesian grid system, which has finite permeability and handled with volume penalization method \cite{angotPenalizationMethodTake1999,deganEffetLanisotropieConvection2002} on the fluid-solid interface. The permeability of solid objects is controlled by $C_{\eta }$, which is kept as $C_{\eta } \ll 1$. The value of $C_{\eta}$ is also related to the error of solution by penalized Navier-Stokes equations. When $C_{\eta}\to 0$, the solution tends to be the exact solution of incompressible Navier-Stokes equations on the no-slip boundary conditions. But a too small $C_{\eta}$ results in large discretization error \cite{nguyenvanyenApproximationLaplaceStokes2014}. 
Hence there exists a range of $C_{\eta}$ to be chosen 
in the calculation. The relation of $C_{\eta}$ and lattice spacing $\Delta x$ of the smallest element in grid system is defined as
\begin{equation}
C_{\eta} = (K_{\eta}^{2}/\nu)\Delta x^{2}
\label{C_eta_def}
\end{equation}
where $K_{\eta}$ is a constant \cite{engelsNumericalSimulationFluid2015} and $\nu$ is normalized kinematic viscosity. In the present paper, all the dragonfly cases keep the constant $K_{\eta}$ as 0.12. The mask function $\chi$ is an indicator of domain and defined as 
\begin{equation}
\chi(\underline{x},t) = \begin{cases}
    0 \quad \mathrm{if}\, \underline{x}\in \Omega_{f}\\
    1 \quad \mathrm{if}\, \underline{x}\in \Omega_{s}
    \end{cases},
\label{chi_def}
\end{equation}
where $\Omega_{f}$ and $\Omega_{s}$ represent fluid and solid domain, respectively. $\mathtt{W} \mathtt{A} \mathtt{B} \mathtt{B} \mathtt{I} \mathtt{T}$ also has great ability in automatic grid refinement, combining with volume penalization method, which can 
handle the CFD simulations of irregular solid objects with high accuracy.

\begin{figure}
\centerline{\includegraphics[width=0.45\textwidth]{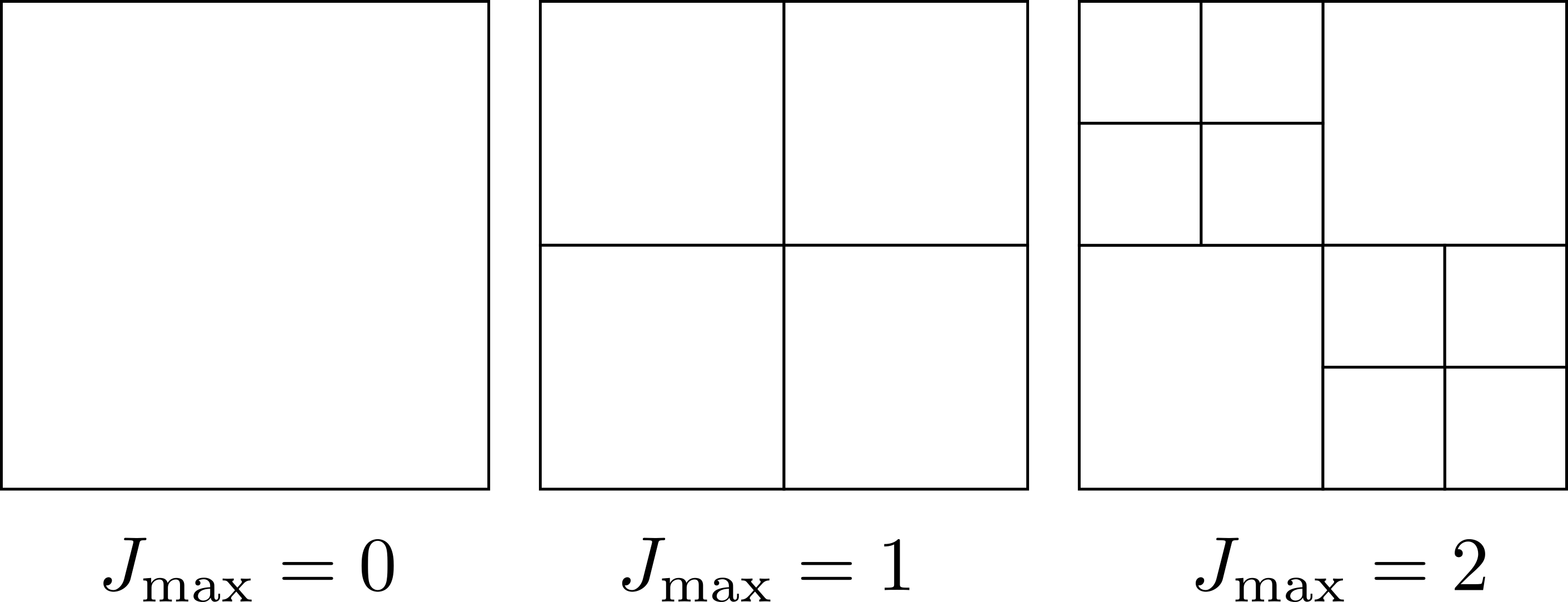}}
\caption{{Schematic diagram of the grid refinement in 2D. Each square is a block containing a number of grid points ($B_s$, not shown for readability). The grid is generated from a single initial block. When $J_\mathrm{max}$ is incremented by one, each block can be refined to four (in 3D eight) new blocks, if the wavelet criterion detects that this block is to be refined. Blocks that need no refinement are kept coarsened where possible, the example shows two blocks on $J=1$ that are not refined when incrementing $J_\mathrm{max}$.}}
\label{fig:Jmax}
\end{figure}

The flight of dragonfly is in much lower Mach number, of which the fluid can be regarded as incompressible flow. We use the 
artificial compressibility method to deal with the incompressible Navier-Stokes equations so as to be free of solving elliptical 
problems. The biorthogonal interpolating wavelet technique is another key feature of $\mathtt{W} \mathtt{A} \mathtt{B} \mathtt{B} \mathtt{I} \mathtt{T}$, 
which is used to tracking the data both in space and scale so as to improve the computational competence in high resolution 
simulations. 
The wavelet-based adaptivity used in $\mathtt{W} \mathtt{A} \mathtt{B} \mathtt{B} \mathtt{I} \mathtt{T}$ 
with MPI parallelization ensures the high efficiency on dragonfly flow field simulations. 
$\Delta x$ in equation \ref{C_eta_def} symbolizes the minimum lattice spacing in the Cartisian grid system, which is controlled by the maximum refinement level $J_\mathrm{max}$ (figure \ref{fig:Jmax}) with
\begin{equation}
\Delta x = \frac{L}{2^{J_\mathrm{max}}B_{s}}
\label{delta_x}
\end{equation}
where $B_{s}$ is the block size and $L$ is the size of the computing domain. 
In this study, all simulations are done on supercomputers equipped with AMD EPYC Milan CPUs. The average cost for a case of $J_\mathrm{max} = 7$ and $\Delta x = 3.906 \times 10^{-3}$ is about 4,000 CPUh on 128 CPU cores. 

\section*{Appendix B. Validation simulations}\label{appendix-B}
\paragraph*{Fruit fly simulations}
{Our code has been extensively validated against other numerical methods \cite{engelsWaveletAdaptiveMethodMultiscale2021} as well as experiments \cite{kolomenskiyAerodynamicPerformanceBristled2020,farisenkovNovelFlightStyle2022}. In this section, we provide an additional validation to further increase confidence in our approach. A central difficulty for validation is that suitable reference data are scarce as we require a high level of accuracy and precise parameter input. Because we are not aware of any data on dragonflies with flat wing models that match those criteria, 
we instead select the case of a hovering fruitfly, initially published in \cite{maedaGroundEffectFruit2013}.  This case is well documented and relevant to the present work, even though dragonflies and fruitflies obviously differ in many aspects. They share moving wings, which pose a significant challenge for computational simulations, and a body. From a computational point of view, they are thus more similar than from a biological perspective.}  

{As is typical in CFD, we normalize the governing equations
using the wing length $R$, frequency $f$ and the density of air, $\varrho_\mathrm{air}$, which yields a dimensionless viscosity of $\nu=1.13 \times 10^{-2}$. The conventional Reynolds number, based on the mean chord length $c_{\mathrm{m}}=0.33$ and the wing tip velocity $u_\mathrm{tip}=4.66$, is $Re=c_{\mathrm{m}} u_\mathrm{tip}/\nu=136$. This is significantly smaller than for a dragonfly but still well in the nonlinear flow regime. Our computational domain is a cube of size $L=6R$, we allow for up to $J_\mathrm{max}=6$ levels of grid refinement resulting in a resolution of $\Delta x = 4.687 \times 10^{-3}$, and set the penalization parameter to $K_\eta=0.22$. For reproducibility, the complete list of numerical parameters including relevant configuration files are included in the supplementary material.}

{Figure \ref{fig:fig5_maeda_verifi} shows the forward force $F_x$ as well as the vertical force $F_z$ , both in the global coordinate system, where $z$ is aligned with gravity. Both show good agreement. The flow field, visualized by the Q-criterion\cite{huntEddiesStreamsConvergence1988} in figure \ref{fig:fig5_maeda_verifi} right, exhibits a complicated vortex pattern, which stresses that despite the difference in Reynolds number, this validation is relevant also for dragonflies.}

\begin{figure}
\centering
\centerline{\includegraphics[width=1.0\textwidth]{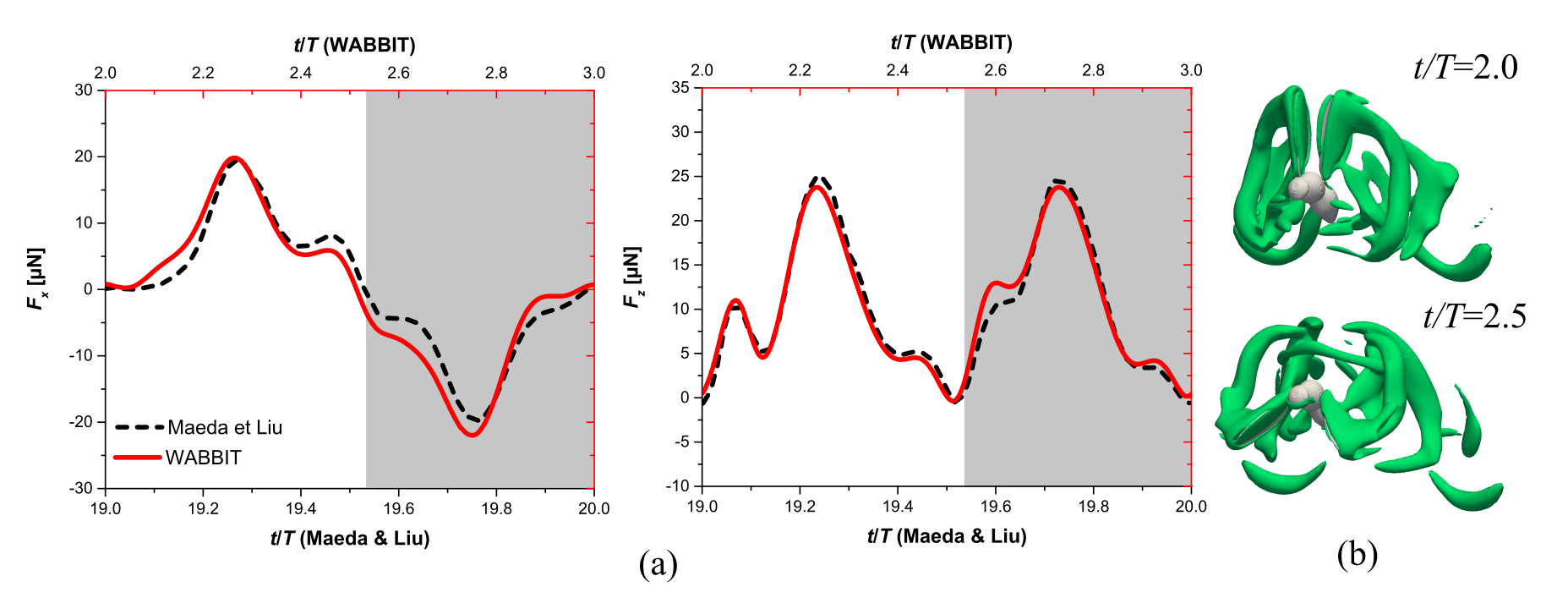}}
\caption{(a) The aerodynamic forces of the fruit fly wing compared with that of \cite{maedaGroundEffectFruit2013}. The grey zones indicate the upstroke of fruit fly. 
The results of Maeda \& Liu \cite{maedaGroundEffectFruit2013} were from 19$\mathrm{^{th}}$ cycle with the bottom x axis, while the stable results of the present study are from 3$\mathrm{^{rd}}$ cycle with the top x axis since we in the third cycle the calculation is converged.; {(b) Snapshots of vortex structures with Q-crition $Q=20$.}}
\label{fig:fig5_maeda_verifi}
\end{figure}

\begin{figure}
\centerline{\includegraphics[width=.40\textwidth]{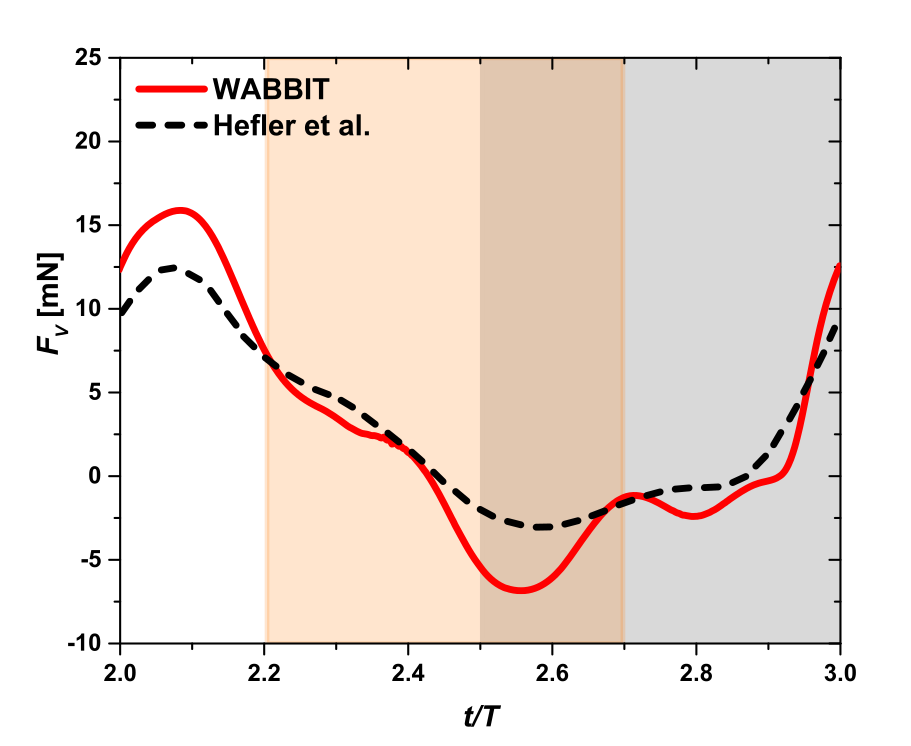}}
\caption{{The total vertical force comparison between the baseline grid in this paper and \cite{heflerAerodynamicPerformanceFreeflying2020}. Note the results reported in \cite{heflerAerodynamicPerformanceFreeflying2020} use a different timeline and have been shifted for comparison. Grey (orange) area indicates forewing (hindwing) upstroke.}}
\label{fig:total_vertical-comparison}
\end{figure}

\begin{figure}
\centering
\begin{subfigure}{.45\textwidth}
    \centering
    \includegraphics[width=\textwidth]{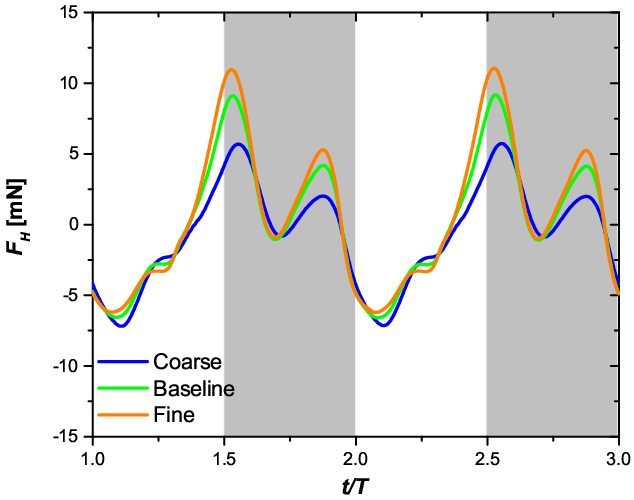}
    \caption{Horizontal force of dragonfly}
\end{subfigure}
\begin{subfigure}{.45\textwidth}
    \centering
    \includegraphics[width=\textwidth]{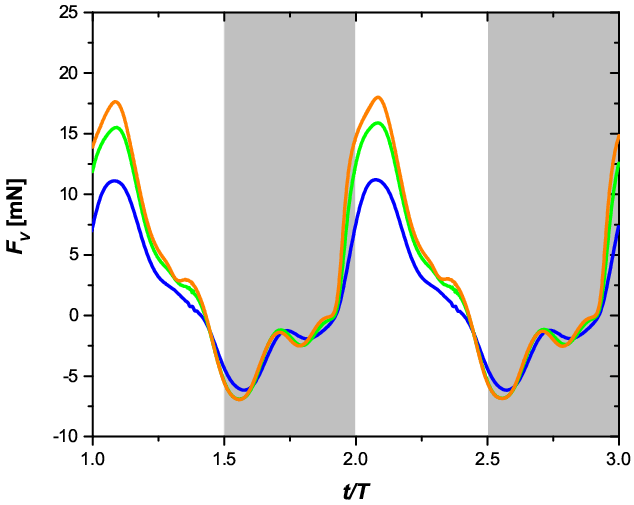}
    \caption{Vertical force of dragonfly}
\end{subfigure}
\caption{The comparison of dragonfly horizontal and vertical forces at three levels of simulations. 
Grey zones indicate the forewing upstrokes.}
\label{fig:fig6_g.i.t._forces}
\end{figure}

\paragraph*{Dragonfly simulations}
Using the numerical wind tunnel defined in the main text, a grid independence test of our dragonfly model is done with three levels of grid, which are listed in detail in table \ref{tab:g.i.t._para} and 
the parameters for dragonfly in forward flight are summarized in table \ref{tab:dragonfly_para}. The average horizontal and vertical forces of the baseline grid $J_\mathrm{max} = 7$ are 0.19mN and 2.50mN, respectively. The calculated horizontal force in \cite{heflerAerodynamicPerformanceFreeflying2020} is 1.15mN and their measured gravity of the dragonfly is 3.28mN. Our vertical force is slightly smaller than the measured data, which can be attributed to the plate wing that we use in the simulations rather than the wing with prescribed twisting. Our horizontal force is closer to 0, which means the dragonfly is more like to be in the ideal stable flight. Figure \ref{fig:total_vertical-comparison} shows the comparison of the total vertical force variation (i.e., 4 wings + body) between our results and the reference \cite{heflerAerodynamicPerformanceFreeflying2020}. The trends and general force of our model agree with the results from \cite{heflerAerodynamicPerformanceFreeflying2020}, which validates the simplified model that we use in this paper.
The results of the grid independence test shown in figure \ref{fig:fig6_g.i.t._forces} show that the aerodynamic forces converge in the third cycle of flapping and the baseline resolution with $J_\mathrm{max} = 7$ is sufficient for the simulations of the dragonfly in the present paper with a good balance of computational resource consumption and result accuracy. 
The resolution of $J_\mathrm{max} = 7$ would correspond, if all blocks were refined to the maximum level, to an equidistant resolution of $2560^3$  grid points. The minimum lattice spacing is $\Delta x = 3.906 \times 10^{-3}$, and the relative wing thickness is $h_{\mathrm{\mathrm{wing}}}/\Delta x = 6.40$. The penalization constant is $K_{\eta}=0.12$ in all present dragonfly simulations.

\begin{table}
\begin{center}
\def~{\hphantom{0}}
\begin{tabular}{|l|l|l|l|l|}\hline
    Level    &  $J_\mathrm{max}$  & $\Delta x$ & $h_{\mathrm{wing}}/\Delta x$  & $C_{\eta}$ \\\hline
    Coarse    &  6  & 7.812 $\times 10^{-3}$ & 3.2 & 4.02 $\times 10^{-3}$ \\\hline
    Baseline    &  7  & 3.906 $\times 10^{-3}$ & 6.4 & 1.00 $\times 10^{-3}$ \\\hline
    Fine    &  8  & 1.953 $\times 10^{-3}$ & 12.8 & 2.51 $\times 10^{-4}$ \\ \hline 
\end{tabular}
\caption{Computational parameters of three levels of simulations in grid independence test for our dragonfly model.}
\label{tab:g.i.t._para}
\end{center}
\end{table}
 
\enlargethispage{20pt}

\section*{Appendix C. Details of the Dragonfly model}\label{appendix-C}
\begin{figure}
\centering
\centerline{\includegraphics[width=0.85\textwidth]{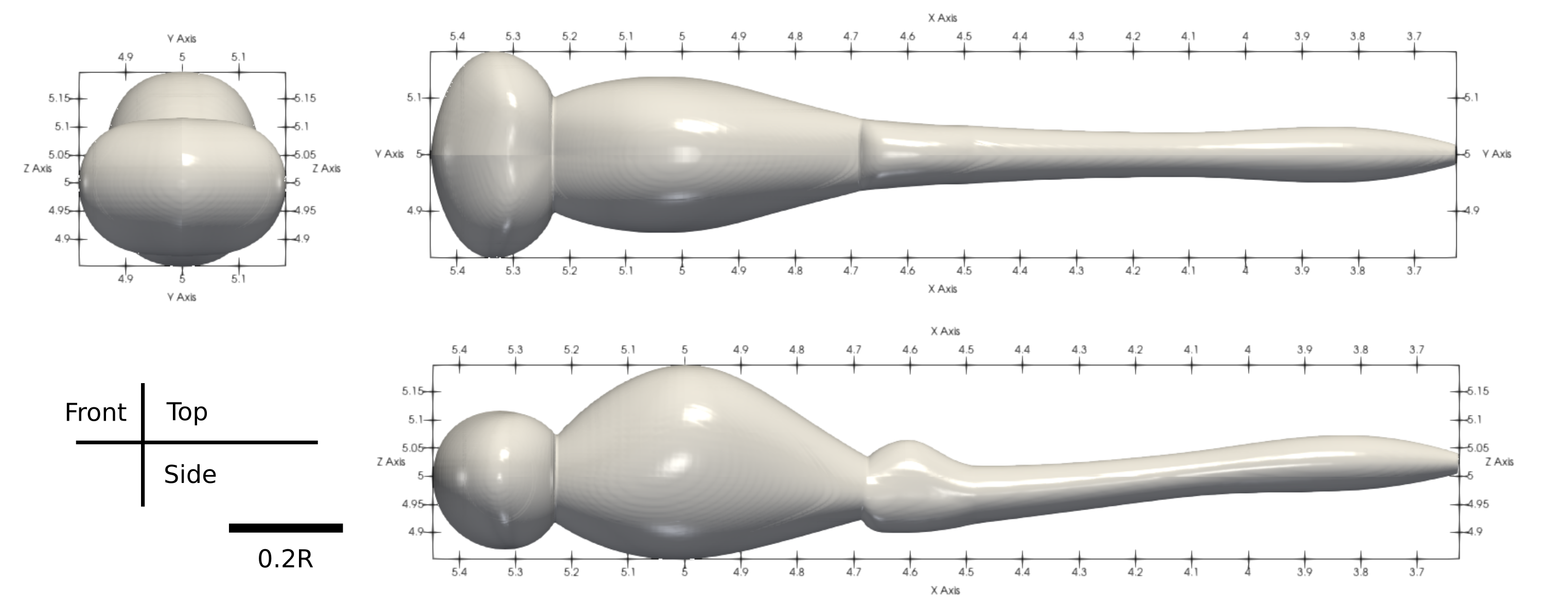}}
\caption{The three views of the dragonfly body used in the simulations.}
\label{fig:dragonfly_body}
\end{figure}

The dragonfly wing shapes are derived from the figures in \cite{heflerAerodynamicPerformanceFreeflying2020}, the wing root locations are $x_{FW,root}=(0.145, \pm 0.2, 0.1)$ and $x_{HW,root}=(-0.0094, \pm 0.2, 0.097)$, respectively. The wing shape is drawn with 41$^\mathrm{th}$ order Fourier transform in polar coordinate system with
\begin{equation}
r = A_{0}/2 + \sum_n A_{n}\mathrm{cos}(n \theta)+B_{n}\mathrm{sin}(n \theta), n \in \{1, 2, 3, \dots, 40\}
\label{delta_x}
\end{equation}
where the range of $\theta$ is $[0, 2\pi)$. The Fourier values for reproducing can be found in the configuration files in supplementary materials.
The dragonfly body model is derived from the real dragonfly with some simplifications. The three views of the dragonfly body are shown in figure \ref{fig:dragonfly_body}.

\section*{Acknowledgments}
{The authors appreciate the help of Dr. Hamed Rajabi at London South Bank University, the support from China Scholarship Council (CSC) and the support of computing resources from MeSU at Sorbonne Université, Paris, France.}

\bibliography{MyLibrary}

\begin{thebibliography}{46}
\newcommand{\enquote}[1]{``#1''}
\providecommand{\natexlab}[1]{#1}
\providecommand{\url}[1]{\texttt{#1}}
\providecommand{\urlprefix}{URL }
\expandafter\ifx\csname urlstyle\endcsname\relax
  \providecommand{\doi}[1]{\discretionary{}{}{}https://doi.org/#1}\else
  \providecommand{\doi}[1]{\discretionary{}{}{}\urlstyle{rm}\url{https://doi.org/#1}}\fi

\bibitem[{Sane(2003)}]{Sane:2003}
Sane, S.~P., \enquote{The aerodynamics of insect flight,} \emph{Journal of
  Experimental Biology}, Vol. 206, No.~23, 2003, pp. 4191--4208.

\bibitem[{Wang(2005)}]{Wang:2005}
Wang, Z.~J., \enquote{Dissecting Insect Flight,} \emph{Annual Review of Fluid
  Mechanics}, Vol.~37, No.~1, 2005, pp. 183--210.
\newblock \doi{10.1146/annurev.fluid.36.050802.121940}.

\bibitem[{Sun(2014)}]{Sun:2014}
Sun, M., \enquote{Insect Flight Dynamics: {{Stability}} and Control,}
  \emph{Reviews of Modern Physics}, Vol.~86, No.~2, 2014, pp. 615--646.
\newblock \doi{10.1103/RevModPhys.86.615}.

\bibitem[{Chin and Lentink(2016)}]{Chin:2016}
Chin, D.~D., and Lentink, D., \enquote{Flapping Wing Aerodynamics: From Insects
  to Vertebrates,} \emph{Journal of Experimental Biology}, Vol. 219, No.~7,
  2016, pp. 920--932.
\newblock \doi{10.1242/jeb.042317}.

\bibitem[{Rajabi et~al.(2020)Rajabi, Dirks, and
  Gorb}]{rajabiInsectWingDamage2020}
Rajabi, H., Dirks, J.-H., and Gorb, S.~N., \enquote{Insect Wing Damage: Causes,
  Consequences and Compensatory Mechanisms,} \emph{Journal of Experimental
  Biology}, Vol. 223, No.~9, 2020, p. jeb215194.
\newblock \doi{10.1242/jeb.215194}.

\bibitem[{Foster and Cartar(2011)}]{fosterWhatCausesWing2011}
Foster, D.~J., and Cartar, R.~V., \enquote{What Causes Wing Wear in Foraging
  Bumble Bees?} \emph{Journal of Experimental Biology}, Vol. 214, No.~11, 2011,
  pp. 1896--1901.
\newblock \doi{10.1242/jeb.051730}.

\bibitem[{Jantzen and Eisner(2008)}]{jantzenHindwingsAreUnnecessary2008}
Jantzen, B., and Eisner, T., \enquote{Hindwings Are Unnecessary for Flight but
  Essential for Execution of Normal Evasive Flight in {{Lepidoptera}},}
  \emph{Proceedings of the National Academy of Sciences}, Vol. 105, No.~43,
  2008, pp. 16636--16640.
\newblock \doi{10.1073/pnas.0807223105}.

\bibitem[{Combes et~al.(2010)Combes, Crall, and
  Mukherjee}]{combesDynamicsAnimalMovement2010}
Combes, S.~A., Crall, J.~D., and Mukherjee, S., \enquote{Dynamics of Animal
  Movement in an Ecological Context: Dragonfly Wing Damage Reduces Flight
  Performance and Predation Success,} \emph{Biology Letters}, Vol.~6, No.~3,
  2010, pp. 426--429.
\newblock \doi{10.1098/rsbl.2009.0915}.

\bibitem[{Rudolf et~al.(2019)Rudolf, Wang, Gorb, and
  Rajabi}]{rudolfFractureResistanceDragonfly2019}
Rudolf, J., Wang, L.-Y., Gorb, S., and Rajabi, H., \enquote{On the Fracture
  Resistance of Dragonfly Wings,} \emph{Journal of the Mechanical Behavior of
  Biomedical Materials}, Vol.~99, 2019, pp. 127--133.
\newblock \doi{10.1016/j.jmbbm.2019.07.009}.

\bibitem[{Rajabi et~al.(2017)Rajabi, Schroeter, Eshghi, and
  Gorb}]{rajabiProbabilityWingDamage2017}
Rajabi, H., Schroeter, V., Eshghi, S., and Gorb, S.~N., \enquote{The
  Probability of the Wing Damage in the Dragonfly {{{\emph{Sympetrum}}}}{\emph{
  Vulgatum}} ({{Anisoptera}}: {{Libellulidae}}): A Field Study,} \emph{Biology
  Open}, 2017, p. bio.027078.
\newblock \doi{10.1242/bio.027078}.

\bibitem[{Engels et~al.(2020)Engels, Wehmann, and
  Lehmann}]{engelsThreedimensionalWingStructure2020}
Engels, T., Wehmann, H.-N., and Lehmann, F.-O., \enquote{Three-Dimensional Wing
  Structure Attenuates Aerodynamic Efficiency in Flapping Fly Wings,}
  \emph{Journal of The Royal Society Interface}, Vol.~17, No. 164, 2020, p.
  20190804.
\newblock \doi{10.1098/rsif.2019.0804}.

\bibitem[{Kihlstr{\"o}m et~al.(2021)Kihlstr{\"o}m, Aiello, Warrant, Sponberg,
  and St{\"o}ckl}]{kihlstromWingDamageAffects2021}
Kihlstr{\"o}m, K., Aiello, B., Warrant, E., Sponberg, S., and St{\"o}ckl, A.,
  \enquote{Wing Damage Affects Flight Kinematics but Not Flower Tracking
  Performance in Hummingbird Hawkmoths,} \emph{Journal of Experimental
  Biology}, Vol. 224, No.~4, 2021, p. jeb236240.
\newblock \doi{10.1242/jeb.236240}.

\bibitem[{Le~Roy et~al.(2019)Le~Roy, Cornette, Llaurens, and
  Debat}]{leroyEffectsNaturalWing2019}
Le~Roy, C., Cornette, R., Llaurens, V., and Debat, V., \enquote{Effects of
  Natural Wing Damage on Flight Performance in {{{\emph{Morpho}}}} Butterflies:
  What Can It Tell Us about Wing Shape Evolution?} \emph{Journal of
  Experimental Biology}, 2019, p. jeb.204057.
\newblock \doi{10.1242/jeb.204057}.

\bibitem[{Meng et~al.(2023)Meng, Liu, Chen, Wu, and
  Chen}]{mengWingKinematicsMeasurement2023}
Meng, X., Liu, X., Chen, Z., Wu, J., and Chen, G., \enquote{Wing Kinematics
  Measurement and Aerodynamics of Hovering Droneflies with Wing Damage,}
  \emph{Bioinspiration \& Biomimetics}, Vol.~18, No.~2, 2023, p. 026013.
\newblock \doi{10.1088/1748-3190/acb97c}.

\bibitem[{Muijres et~al.(2017)Muijres, Iwasaki, Elzinga, Melis, and
  Dickinson}]{muijresFliesCompensateUnilateral2017}
Muijres, F.~T., Iwasaki, N.~A., Elzinga, M.~J., Melis, J.~M., and Dickinson,
  M.~H., \enquote{Flies Compensate for Unilateral Wing Damage through Modular
  Adjustments of Wing and Body Kinematics,} \emph{Interface Focus}, Vol.~7,
  No.~1, 2017, p. 20160103.
\newblock \doi{10.1098/rsfs.2016.0103}.

\bibitem[{Lyu et~al.(2020)Lyu, Zhu, and Sun}]{lyuWingKinematicAerodynamic2020}
Lyu, Y.~Z., Zhu, H.~J., and Sun, M., \enquote{Wing Kinematic and Aerodynamic
  Compensations for Unilateral Wing Damage in a Small Phorid Fly,}
  \emph{Physical Review E}, Vol. 101, No.~1, 2020, p. 012412.
\newblock \doi{10.1103/PhysRevE.101.012412}.

\bibitem[{Hsieh et~al.(2010)Hsieh, Kung, Chang, and
  Chu}]{hsiehUnsteadyAerodynamicsDragonfly2010}
Hsieh, C.-T., Kung, C.-F., Chang, C.~C., and Chu, C.-C., \enquote{Unsteady
  Aerodynamics of Dragonfly Using a Simple Wing--Wing Model from the
  Perspective of a Force Decomposition,} \emph{Journal of Fluid Mechanics},
  Vol. 663, 2010, pp. 233--252.
\newblock \doi{10.1017/S0022112010003484}.

\bibitem[{Hu and Deng(2014)}]{huAerodynamicInteractionForewing2014}
Hu, Z., and Deng, X.-Y., \enquote{Aerodynamic Interaction between Forewing and
  Hindwing of a Hovering Dragonfly,} \emph{Acta Mechanica Sinica}, Vol.~30,
  No.~6, 2014, pp. 787--799.
\newblock \doi{10.1007/s10409-014-0118-6}.

\bibitem[{{Weis-Fogh}(1973)}]{weis-foghQuickEstimatesFlight1973}
{Weis-Fogh}, T., \enquote{Quick {{Estimates}} of {{Flight Fitness}} in
  {{Hovering Animals}}, {{Including Novel Mechanisms}} for {{Lift
  Production}},} \emph{Journal of Experimental Biology}, Vol.~59, No.~1, 1973,
  pp. 169--230.
\newblock \doi{10.1242/jeb.59.1.169}.

\bibitem[{Saffman and Sheffield(1977)}]{saffmanFlowWingAttached1977}
Saffman, P.~G., and Sheffield, J.~S., \enquote{Flow over a {{Wing}} with an
  {{Attached Free Vortex}},} \emph{Studies in Applied Mathematics}, Vol.~57,
  No.~2, 1977, pp. 107--117.
\newblock \doi{10.1002/sapm1977572107}.

\bibitem[{{van Veen} et~al.(2022){van Veen}, {van Leeuwen}, {van Oudheusden},
  and Muijres}]{vanveenUnsteadyAerodynamicsInsect2022}
{van Veen}, W.~G., {van Leeuwen}, J.~L., {van Oudheusden}, B.~W., and Muijres,
  F.~T., \enquote{The Unsteady Aerodynamics of Insect Wings with Rotational
  Stroke Accelerations, a Systematic Numerical Study,} \emph{Journal of Fluid
  Mechanics}, Vol. 936, 2022, p.~A3.
\newblock \doi{10.1017/jfm.2022.31}.

\bibitem[{Dickinson et~al.(1999)Dickinson, Lehmann, and
  Sane}]{dickinsonWingRotationAerodynamic1999}
Dickinson, M.~H., Lehmann, F.-O., and Sane, S.~P., \enquote{Wing {{Rotation}}
  and the {{Aerodynamic Basis}} of {{Insect Flight}},} \emph{Science}, Vol.
  284, No. 5422, 1999, pp. 1954--1960.
\newblock \doi{10.1126/science.284.5422.1954}.

\bibitem[{Srygley and
  Thomas(2002)}]{srygleyUnconventionalLiftgeneratingMechanisms2002}
Srygley, R.~B., and Thomas, A. L.~R., \enquote{Unconventional Lift-Generating
  Mechanisms in Free-Flying Butterflies,} \emph{Nature}, Vol. 420, No. 6916,
  2002, pp. 660--664.
\newblock \doi{10.1038/nature01223}.

\bibitem[{Gao and Lu(2008)}]{gaoInsectNormalHovering2008}
Gao, T., and Lu, X.-Y., \enquote{Insect Normal Hovering Flight in Ground
  Effect,} \emph{Physics of Fluids}, Vol.~20, No.~8, 2008, p. 087101.
\newblock \doi{10.1063/1.2958318}.

\bibitem[{Kim and
  Kim(2011)}]{kimComputationalInvestigationThreedimensional2011}
Kim, J.-H., and Kim, C., \enquote{Computational {{Investigation}} of
  {{Three-dimensional Unsteady Flowfield Characteristics}} around {{Insects}}'
  {{Flapping Flight}},} \emph{AIAA Journal}, Vol.~49, No.~5, 2011, pp.
  953--968.
\newblock \doi{10.2514/1.J050485}.

\bibitem[{Engels et~al.(2016{\natexlab{a}})Engels, Kolomenskiy, Schneider,
  Lehmann, and Sesterhenn}]{engelsBumblebeeFlightHeavy2016}
Engels, T., Kolomenskiy, D., Schneider, K., Lehmann, F.-O., and Sesterhenn, J.,
  \enquote{Bumblebee {{Flight}} in {{Heavy Turbulence}},} \emph{Physical Review
  Letters}, Vol. 116, No.~2, 2016{\natexlab{a}}, p. 028103.
\newblock \doi{10.1103/PhysRevLett.116.028103}.

\bibitem[{Engels et~al.(2019)Engels, Kolomenskiy, Schneider, Farge, Lehmann,
  and Sesterhenn}]{engelsImpactTurbulenceFlying2019}
Engels, T., Kolomenskiy, D., Schneider, K., Farge, M., Lehmann, F.-O., and
  Sesterhenn, J., \enquote{Impact of Turbulence on Flying Insects in Tethered
  and Free Flight: {{High-resolution}} Numerical Experiments,} \emph{Physical
  Review Fluids}, Vol.~4, No.~1, 2019, p. 013103.
\newblock \doi{10.1103/PhysRevFluids.4.013103}.

\bibitem[{Bhat et~al.(2020)Bhat, Zhao, Sheridan, Hourigan, and
  Thompson}]{bhatEffectsFlappingmotionProfiles2020}
Bhat, S.~S., Zhao, J., Sheridan, J., Hourigan, K., and Thompson, M.~C.,
  \enquote{Effects of Flapping-Motion Profiles on Insect-Wing Aerodynamics,}
  \emph{Journal of Fluid Mechanics}, Vol. 884, 2020, p.~A8.
\newblock \doi{10.1017/jfm.2019.929}.

\bibitem[{Mittal and Iaccarino(2005)}]{mittalIMMERSEDBOUNDARYMETHODS2005}
Mittal, R., and Iaccarino, G., \enquote{{{IMMERSED BOUNDARY METHODS}},}
  \emph{Annual Review of Fluid Mechanics}, Vol.~37, No. Volume 37, 2005, 2005,
  pp. 239--261.
\newblock \doi{10.1146/annurev.fluid.37.061903.175743}.

\bibitem[{Schneider(2015)}]{schneiderImmersedBoundaryMethods2015}
Schneider, K., \enquote{Immersed Boundary Methods for Numerical Simulation of
  Confined Fluid and Plasma Turbulence in Complex Geometries: A Review,}
  \emph{Journal of Plasma Physics}, Vol.~81, No.~6, 2015, p. 435810601.
\newblock \doi{10.1017/S0022377815000598}.

\bibitem[{Hefler et~al.(2020)Hefler, Noda, Qiu, and
  Shyy}]{heflerAerodynamicPerformanceFreeflying2020}
Hefler, C., Noda, R., Qiu, H.~H., and Shyy, W., \enquote{Aerodynamic
  Performance of a Free-Flying Dragonfly---{{A}} Span-Resolved Investigation,}
  \emph{Physics of Fluids}, Vol.~32, No.~4, 2020, p. 041903.
\newblock \doi{10.1063/1.5145199}.

\bibitem[{Engels et~al.(2016{\natexlab{b}})Engels, Kolomenskiy, Schneider, and
  Sesterhenn}]{engelsFluSINovelParallel2016}
Engels, T., Kolomenskiy, D., Schneider, K., and Sesterhenn, J.,
  \enquote{{{FluSI}}: {{A Novel Parallel Simulation Tool}} for {{Flapping
  Insect Flight Using}} a {{Fourier Method}} with {{Volume Penalization}},}
  \emph{SIAM Journal on Scientific Computing}, Vol.~38, No.~5,
  2016{\natexlab{b}}, pp. S3--S24.
\newblock \doi{10.1137/15M1026006}.

\bibitem[{Engels et~al.(2021)Engels, Schneider, Reiss, and
  Farge}]{engelsWaveletAdaptiveMethodMultiscale2021}
Engels, T., Schneider, K., Reiss, J., and Farge, M., \enquote{A
  {{Wavelet-Adaptive Method}} for {{Multiscale Simulation}} of {{Turbulent
  Flows}} in {{Flying Insects}},} \emph{Communications in Computational
  Physics}, Vol.~30, No.~4, 2021, pp. 1118--1149.
\newblock \doi{10.4208/cicp.OA-2020-0246}.

\bibitem[{Angot et~al.(1999)Angot, Bruneau, and
  Fabrie}]{angotPenalizationMethodTake1999}
Angot, P., Bruneau, C.-H., and Fabrie, P., \enquote{A Penalization Method to
  Take into Account Obstacles in Incompressible Viscous Flows,}
  \emph{Numerische Mathematik}, Vol.~81, No.~4, 1999, pp. 497--520.
\newblock \doi{10.1007/s002110050401}.

\bibitem[{Ohwada and Asinari(2010)}]{ohwadaArtificialCompressibilityMethod2010}
Ohwada, T., and Asinari, P., \enquote{Artificial Compressibility Method
  Revisited: {{Asymptotic}} Numerical Method for Incompressible
  {{Navier}}--{{Stokes}} Equations,} \emph{Journal of Computational Physics},
  Vol. 229, No.~5, 2010, pp. 1698--1723.
\newblock \doi{10.1016/j.jcp.2009.11.003}.

\bibitem[{Hunt et~al.(1988)Hunt, Wray, and
  Moin}]{huntEddiesStreamsConvergence1988}
Hunt, J. C.~R., Wray, A., and Moin, P., \enquote{Eddies, Streams, and
  Convergence Zones in Turbulent Flows,} \emph{Proceedings of the {{Summer
  Program}}}, Center for Turbulence Research, Stanford University, 1988, pp.
  193--208.

\bibitem[{Van Den~Berg and Ellington(1997)}]{vandenbergVortexWakeHovering1997}
Van Den~Berg, C., and Ellington, C.~P., \enquote{The Vortex Wake of a
  `Hovering' Model Hawkmoth,} \emph{Philosophical Transactions of the Royal
  Society of London. Series B: Biological Sciences}, Vol. 352, No. 1351, 1997,
  pp. 317--328.
\newblock \doi{10.1098/rstb.1997.0023}.

\bibitem[{Birch et~al.(2004)Birch, Dickson, and
  Dickinson}]{birchForceProductionFlow2004}
Birch, J.~M., Dickson, W.~B., and Dickinson, M.~H., \enquote{Force Production
  and Flow Structure of the Leading Edge Vortex on Flapping Wings at High and
  Low {{Reynolds}} Numbers,} \emph{Journal of Experimental Biology}, Vol. 207,
  No.~7, 2004, pp. 1063--1072.
\newblock \doi{10.1242/jeb.00848}.

\bibitem[{{Bode-Oke} et~al.(2018){Bode-Oke}, Zeyghami, and
  Dong}]{bode-okeFlyingReverseKinematics2018}
{Bode-Oke}, A.~T., Zeyghami, S., and Dong, H., \enquote{Flying in Reverse:
  Kinematics and Aerodynamics of a Dragonfly in Backward Free Flight,}
  \emph{Journal of The Royal Society Interface}, Vol.~15, No. 143, 2018, p.
  20180102.
\newblock \doi{10.1098/rsif.2018.0102}.

\bibitem[{Birch and Dickinson(2001)}]{birchSpanwiseFlowAttachment2001}
Birch, J.~M., and Dickinson, M.~H., \enquote{Spanwise Flow and the Attachment
  of the Leading-Edge Vortex on Insect Wings,} \emph{Nature}, Vol. 412, 2001,
  pp. 729--733.
\newblock \doi{doi.org/10.1038/35089071}.

\bibitem[{Degan(2002)}]{deganEffetLanisotropieConvection2002}
Degan, G., \enquote{Effet de l'anisotropie Sur La Convection Naturelle
  Engendr{\'e}e Par Une Source Thermique sans Un Milieu Poreux,} \emph{Journal
  de la Recherche Scientifique de l'Universit{\'e} de Lom{\'e}}, Vol.~6, No.~1,
  2002, pp. 131--136.

\bibitem[{Nguyen Van~Yen et~al.(2014)Nguyen Van~Yen, Kolomenskiy, and
  Schneider}]{nguyenvanyenApproximationLaplaceStokes2014}
Nguyen Van~Yen, R., Kolomenskiy, D., and Schneider, K., \enquote{Approximation
  of the {{Laplace}} and {{Stokes}} Operators with {{Dirichlet}} Boundary
  Conditions through Volume Penalization: A Spectral Viewpoint,}
  \emph{Numerische Mathematik}, Vol. 128, No.~2, 2014, pp. 301--338.
\newblock \doi{10.1007/s00211-014-0610-8}.

\bibitem[{Engels et~al.(2015)Engels, Kolomenskiy, Schneider, and
  Sesterhenn}]{engelsNumericalSimulationFluid2015}
Engels, T., Kolomenskiy, D., Schneider, K., and Sesterhenn, J.,
  \enquote{Numerical Simulation of Fluid--Structure Interaction with the Volume
  Penalization Method,} \emph{Journal of Computational Physics}, Vol. 281,
  2015, pp. 96--115.
\newblock \doi{10.1016/j.jcp.2014.10.005}.

\bibitem[{Kolomenskiy et~al.(2020)Kolomenskiy, Farisenkov, Engels, Lapina,
  Petrov, Lehmann, Onishi, Liu, and
  Polilov}]{kolomenskiyAerodynamicPerformanceBristled2020}
Kolomenskiy, D., Farisenkov, S., Engels, T., Lapina, N., Petrov, P., Lehmann,
  F.-O., Onishi, R., Liu, H., and Polilov, A., \enquote{Aerodynamic Performance
  of a Bristled Wing of a Very Small Insect: {{Dynamically}} Scaled Model
  Experiments and Computational Fluid Dynamics Simulations Using a Revolving
  Wing Model,} \emph{Experiments in Fluids}, Vol.~61, No.~9, 2020, p. 194.
\newblock \doi{10.1007/s00348-020-03027-0}.

\bibitem[{Farisenkov et~al.(2022)Farisenkov, Kolomenskiy, Petrov, Engels,
  Lapina, Lehmann, Onishi, Liu, and Polilov}]{farisenkovNovelFlightStyle2022}
Farisenkov, S.~E., Kolomenskiy, D., Petrov, P.~N., Engels, T., Lapina, N.~A.,
  Lehmann, F.-O., Onishi, R., Liu, H., and Polilov, A.~A., \enquote{Novel
  Flight Style and Light Wings Boost Flight Performance of Tiny Beetles,}
  \emph{Nature}, Vol. 602, No. 7895, 2022, pp. 96--100.
\newblock \doi{10.1038/s41586-021-04303-7}.

\bibitem[{Maeda and Liu(2013)}]{maedaGroundEffectFruit2013}
Maeda, M., and Liu, H., \enquote{Ground {{Effect}} in {{Fruit Fly Hovering}}:
  {{A Three-Dimensional Computational Study}},} \emph{Journal of Biomechanical
  Science and Engineering}, Vol.~8, No.~4, 2013, pp. 344--355.
\newblock \doi{10.1299/jbse.8.344}.

\end{thebibliography}

\end{document}


\maketitle

\section{Cycle-averaged forces and power of the different models}

The horizontal force ($\Bar{F_H}$), vertical force ($\Bar{F_V}$) and power ($\Bar{p}$) are calculated in average of one single damaged wing in 3$\mathrm{rd}$ cycle, and $\mathsf{IntactFW}$/$\mathsf{IntactHW}$ are the values of the fore-/hindwing in the intact case, respectively.

\begin{table}[H]
\begin{center}
\def~{\hphantom{0}}
\begin{tabular}{llllllll}
    Case  &  $\Bar{F_H}$  & $\Bar{F_V}$ & $\Bar{p}$ & Case  &  $\Bar{F_H}$  & $\Bar{F_V}$ & $\Bar{p}$ \\
    $\mathsf{IntactFW}$   &  -0.1028mN  &  0.8819mN & 0.00305W & $\mathsf{IntactHW}$   &  0.2376mN  & 0.3704mN & 0.00337W \\
    $\mathsf{FW1}$      &  -0.1154mN  & 0.8281mN & 0.00301W & $\mathsf{HW1}$      &  0.2298mN  & 0.3064mN & 0.00300W \\
    $\mathsf{FW2}$      &  -0.1249mN  & 0.7865mN & 0.00261W & $\mathsf{HW2}$      &  0.1997mN  & 0.2393mN & 0.00245W\\
    $\mathsf{FW3}$      &  -0.1443mN  & 0.7194mN & 0.00231W  & $\mathsf{HW3}$      &  0.1502mN  & 0.1473mN & 0.00179W\\
    $\mathsf{FW4}$      &  -0.1444mN  & 0.6407mN & 0.00195W & $\mathsf{HW4}$      &  0.1217mN  & 0.1211mN & 0.00154W \\
    $\mathsf{FW5}$      &  -0.1398mN  & 0.1777mN & 0.00030W & $\mathsf{HW5}$      &  -0.0123mN  & 0.0253mN & 0.00026W\\
\end{tabular}
\label{tab:physical_forces_power}
\end{center}
\end{table}\textbf{}

\section{Damaged wing force coefficients for all cases}

\begin{figure}[H]
\centerline{\includegraphics[width=0.95\textwidth]{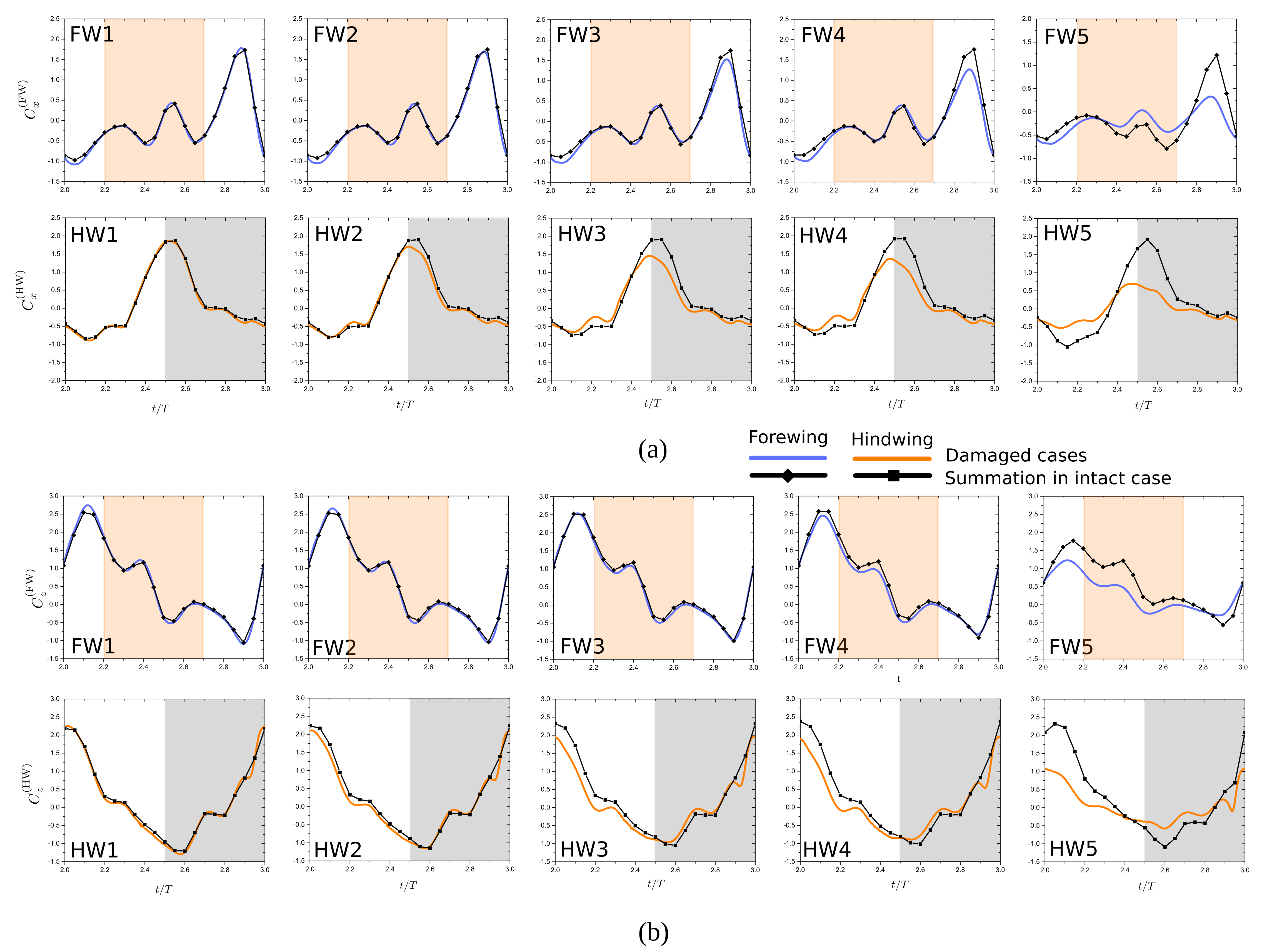}}
\caption{Damaged wing force coefficients for all cases (colored lines) vs. an hypothetical force coefficient computed from the intact wing flow field but where the force integration is performed only over the area that corresponds to each damaged case (black lines, see text). Grey (orange) areas indicate forewing (hindwing) upstroke. (a) Horizontal force coefficient comparisons. (b) Vertical force coefficient comparisons.}
\label{fig:damagedvsintact}
\end{figure}

\section{Video of the interaction wake structure in the intact and $\mathsf{HW5}$}

See video intact-interaction\_wake.webm and fw5-interaction\_wake.webm.

\section{List of numerical parameters including relevant configuration files}

See in compressed file Configurations.zip or the link OSF.IO/2JAU6.